\documentclass[twocolumn,10pt]{IEEEtran}
\usepackage{acronym}
\usepackage{cite}
\usepackage{graphicx}
\usepackage{amsmath}
\usepackage{amssymb}
\usepackage{booktabs}
\usepackage{threeparttable}
\usepackage{algorithm}
\usepackage{xcolor}
\usepackage{array,multirow}
\usepackage{tabularx} 
\usepackage{dblfloatfix} 
\usepackage{bm}
\usepackage{hyperref}

\IEEEaftertitletext{\vspace{-2\baselineskip}}

\newcommand{\yt}{\mathbf{y}} 
\newcommand{\yf}{\tilde{\mathbf{y}}} 

\newcommand{\Yf}{\tilde{\mathbf{Y}}}
\newcommand{\Yt}{{\mathbf{Y}}}

\newcommand{\st}{\mathbf{s}}
\newcommand{\sfd}{\tilde{\mathbf{s}}}

\newcommand{\xt}{\mathbf{x}}
\newcommand{\Xt}{\mathbf{X}}
\newcommand{\xf}{\tilde{\mathbf{x}}}

\newcommand{\zt}{\mathbf{z}}
\newcommand{\Zt}{\mathbf{Z}}
\newcommand{\zf}{\tilde{\mathbf{z}}}

\newcommand{\Rt}{\mathbf{R}}
\newcommand{\rt}{\mathbf{r}}
\newcommand{\mrt}{\mathbf{x}_R}

\newcommand{\Hf}{\tilde{\mathbf{H}}}

\newcommand{\hfi}{\tilde{h}}
\newcommand{\Ht}{\mathbf{H}}

\newcommand{\A}{\mathbf{A}}
\newcommand{\av}{\mathbf{a}}

\newcommand{\C}{\mathbf{C}}

\newcommand{\cv}{\mathbf{c}}
\newcommand{\D}{\mathbf{D}}
\newcommand{\df}{\mathbf{d}}

\newcommand{\F}{\mathbf{F}}
\newcommand{\f}{\mathbf{f}}

\newcommand{\I}{\mathbf{I}}
\newcommand{\Pc}{\mathbf{P}}

\newcommand{\V}{\mathbf{V}}

\newcommand{\W}{\mathbf{W}}
\newcommand{\w}{\mathbf{w}}

\newcommand{\bpsi}{\bm{\psi}}

\newcommand{\Hfps}{\bm{B}}
\newcommand{\hfps}{\bm{\beta}}

\DeclareMathOperator*{\argmax}{argmax} 

\begin{document}

\title{An Overview of Signal Processing Techniques for Joint Communication and Radar Sensing}

\author{J. Andrew Zhang,~\IEEEmembership{Senior~Member,~IEEE}, Fan Liu,~\IEEEmembership{Member,~IEEE},\\ Christos Masouros,~\IEEEmembership{Senior~Member,~IEEE}, Robert W. Heath Jr.,~\IEEEmembership{Fellow,~IEEE}, \\Zhiyong Feng,~\IEEEmembership{Senior~Member,~IEEE}, Le Zheng,~\IEEEmembership{Member,~IEEE}, Athina Petropulu,~\IEEEmembership{Fellow,~IEEE}\\
\thanks{J. A. Zhang is with the University of Technology Sydney, Australia. Email: Andrew.Zhang@uts.edu.au; F. Liu, the corresponding author, is with Southern University of Science and Technology, China. Email: liuf6@sustech.edu.cn; C. Masouros is with University College London, UK. Email: c.masouros@ucl.ac.uk; R. W. Heath Jr. is with the North Carolina State University, USA. Email: rwheathjr@ncsu.edu; Z. Feng is with Beijing University of Posts and Telecommunications, China. Email: fengzy@bupt.edu.cn; L. Zheng is with the Aptiv Company, USA. Email: le.2.zheng@aptiv.com; A. Petropulu is with The State University of New Jersey, USA. Email: athinap@soe.rutgers.edu.}
}

\maketitle

\begin{abstract}
Joint communication and radar sensing (JCR) represents an emerging research field aiming to integrate the above two functionalities into a single system, sharing a majority of hardware and signal processing modules and, in a typical case, sharing a single transmitted signal. It is recognised as a key approach in significantly improving spectrum efficiency, reducing device size, cost and power consumption, and improving performance thanks to potential close cooperation of the two functions. Advanced signal processing techniques are critical for making the integration efficient, from transmission signal design to receiver processing. This paper provides a comprehensive overview of JCR systems from the signal processing perspective, with a focus on state-of-the-art. A balanced coverage on both transmitter and receiver is provided for three types of JCR systems, communication-centric, radar-centric, and joint design and optimization.
\end{abstract}


\section{Introduction}\label{sec:intro}

\subsection{Background}
Wireless communication and radar sensing (C\&R) have been advancing in parallel yet with limited intersections for decades.  They share many commonalities in terms of signal processing algorithms, devices and, to a certain extent, system architecture. They can also potentially share the spectrum. These have recently motivated significant research interests in the coexistence, cooperation, and joint design (or co-design) of the two systems \cite{RN32, Han13, RN15, Hassanien16, 7465731,8828016,liu2019joint,Hassanien_2019, 8828030,Ma20,Feng20,Zhangpmn20,Ali20}. 

The coexistence of C\&R systems has been extensively studied in the past decade, with a focus on developing efficient interference management techniques so that the two individually deployed systems can operate smoothly without interfering with each other \cite{RN15,li2016optimum,Li17,Liu18mumimo,8828016,8835700,7898445}. Although C\&R systems may be co-located and even physically integrated, they transmit two different signals overlapped in time and/or frequency domains. They operate cooperatively to minimize interference to each other. Great research efforts have been devoted to mutual interference cancellation in this case, using, for example, beamforming (BF) design in \cite{7898445} and cooperative spectrum sharing in \cite{8835700}. However, effective interference cancellation typically has stringent requirements on the mobility of nodes and information exchange between them. The spectral efficiency improvement is hence limited in practice.

The joint design of colocated C\&R systems considers the integration of these two functions in one system. The initial concept may be traced back to 1960s \cite{mealey1963method}, and the research had been primarily on multimode or multi-function military radars until 2010s. Recently, we are witnessing a booming interest from both academia and industry on the joint system, thanks to its great potentials in emerging defence applications and more recently in a range of Smart Cities applications such as intelligent vehicular networks \cite{Kumari17, 8057284} and more general, the Internet of Things (IoT) \cite{7465731}. Where there is a recognised congestion of sensors and transceivers, integrating the two systems in one can achieve reduced device size, power consumption, and cost, lead to more efficient radio spectrum usage, and is expected to significantly expand the capabilities and performance of both communication and radar sensing.  

The integration may be classified into the following two classes: (1) The two functions are physically integrated in one system, but they use two sets of dedicated hardware components, and/or  two different waveforms, superimposed or separated in time, frequency or spatial domains \cite{Han13}; (2) The two functions are more firmly integrated by sharing the majority of hardware components and are delivered by the same waveform, which  is  designed to optimize both  communication and radar performance \cite{liu2019joint,Hassanien_2019, Zhangpmn20}. The first class represents  a loose integration and may only achieve limited benefits such as reduction in signalling overheads and removal of interference \cite{Zhangpmn20}. The second class is a step change beyond collocation, and is the main focus of this paper.  

Based on the design priorities and the underlying signal and systems,  current joint C\&R systems may be classified into the following three categories:
\begin{itemize}
	\item \textit{Communication-centric design}. In this class,  radar sensing is an add on to a  communication system, where the design priority is  on communications. The aim of such design is to exploit communication waveform to extract radar information through target echoes. Enhancements to hardware and algorithms are required to support radar sensing. Possible enhancements to communication standards may be introduced to enable better reuse of the communication waveform for radar sensing purposes. In this design, the communication performance can be largely unaffected, however, the sensing performance may be scenario-dependent and difficult-to-tune;
	\item \textit{Radar-centric design}. Conversely, such approaches aim at modulating or introducing information signalling in known radar waveform. Since the radar signalling remains largely unaltered, the resulting approaches benefit from a near optimal radar performance. The main drawback of such approaches is the limited data rates achieved. Some performance loss may be tolerated by the radar to  enable better communication functionality; and 
	\item \textit{Joint design and Optimization}. This class encompasses systems that are jointly designed from the start, to offer a tunable trade-off between C\&R performance. Such systems may not be limited by any of the existing communication or radar standards. 	
\end{itemize}
Owing to the significant differences between traditional C\&R systems, the design problems in these three categories are quite different. In the first two categories, the design and research focus is typically on how to realize the secondary radar (communication) function based on the signal formats of the primary communication (radar) system, in a way that does not significantly affect the primary system. The last category considers the joint design and optimization of the signal waveform, system, and network architecture, with a flexible tradeoff achievable between C\&R. 

These categories of joint systems have been receiving strong and growing research interest, with results having been reported under various names, such as Radar-communications (RadCom) \cite{RN32}, joint radar (and) communications (JRC) \cite{8828030,liu2019joint,Feng20}, joint communication (and) radar (JCR) \cite{Kumari17, Kumari20}, joint communication and radar/radio sensing (JCAS) \cite{8550811,8827589}, and dual-function(al) radar communications (DFRC) \cite{Hassanien16, Liu18, Ma20,xu2020dfrc}. The first three typically refer to a general joint system and can be used interchangeably. Sometimes JRC and JCR are used to differentiate between radar-centric and communication-centric designs. The term JCAS is introduced to stress the evolution of radar towards more general radio sensing applications of communication-centric joint systems. These sensing applications go beyond localization, tracking and object recognition of traditional radar functions, such as human behaviour recognition and atmosphere monitoring using radio signals \cite{7465731}. DFRC is specifically used for joint systems with a shared, single transmitted waveform. It has been widely used for radar-centric joint systems \cite{Hassanien16,Hassanien_2019}, and has also been recently used for communication-centric systems \cite{Liu18, Ma20}. In the rest of this paper, we  use the term  JCR to refer to  a general joint system, and use the term {DFRC} to refer to joint systems with a  shared single transmit waveform.  
	
\subsection{Contributions and Structure of this Paper}
There exist some excellent overview articles on JCR, particularly on system and signal modelling. For example, \cite{RN32} provides a foundational review of signal models for basic single carrier and multicarrier JCR systems; \cite{RN15,8828016,Feng20} cover C\&R systems  from coexistence, cooperation, to co-design; \cite{Hassanien16,Hassanien_2019} survey radar-centric DFRC systems, with a focus on signal embedding; \cite{liu2019joint} delivers an excellent review on the evolution of JCR systems and their potential applications; \cite{Zhangpmn20} provides an overview on mobile network JCR systems; \cite{8828030,Ma20,Ali20}   overview millimeter-wave JCR systems for automotive networks, providing  detailed signal models. However, the topic of receiver signal processing, which is critical for the viability and performance of the JCR systems, has not been adequately covered by the aforementioned literature.

The objective of this paper is to provide a comprehensive overview on JCR from the signal processing perspective, with balanced coverage on both transmitter and receiver. For each of the first two categories of JCR designs, we review main systems and typical signal models, and then discuss critical signal processing techniques at the receiver. For the third category, we provide a detailed overview of joint design and optimization techniques. To reduce the overlap with existing articles, we also focus on more recent technologies. The detailed structure of the rest of this paper is as follows.
\begin{itemize}
	\item Section \ref{sec-system} describes the signal and channel models of typical C\&R systems, separately. Via comparing these models, we disclose the connections and differences between C\&R, which are very important for the understanding and development of JCR technologies;
	\item Section \ref{sec-ccd} provides a review on communication-centric JCR technologies. We first introduce two main JCR systems based on 802.11ad and mobile networks, and then discuss signal processing technologies on sensing parameter estimation, resolution of clock asynchrony, and sensing assisted communications;
	\item Section \ref{sec-rcd} reviews radar-centric JCR technologies. Well-known information embedding technologies are first summarized, and the recent promising technology of index modulation (IM) is elaborated, with reference to several emerging radar systems. We then describe signal reception and processing techniques for communications, including information demodulation and channel estimation for IM. Codebook design for IM is also discussed;
	\item A detailed review of joint design and optimization technologies for JCR is provided in Section \ref{sec-waveform}. These technologies include waveform optimization via spatial precoding, multibeam optimization for analog array, and signal optimization in other domains. With these technologies, balanced performance between C\&R can be achieved as desired; and
	\item Concluding remarks are provided in  Section \ref{sec-final}.  
\end{itemize}

\textbf{Notations:} $\mathbb{C}$ denotes the set of complex numbers. $(\cdot)^H$, $(\cdot)^*$, $(\cdot)^T$ denote the Hermitian transpose, conjugate, and transpose of a matrix or vector. $n!$ denotes $n$ factorial and $C_k^n=n!/(k!(n-k)!)$ denotes the binomial coefficient. $\lfloor x\rfloor$ rounds towards negative infinity. $\{x_m\}, m=1,\ldots, M$ denotes a column vector with elements $x_1, \ldots, x_M$.

\section{System and Signal Models}\label{sec-system}

\begin{figure}[t]
	\centering
	\includegraphics[width=1\linewidth]{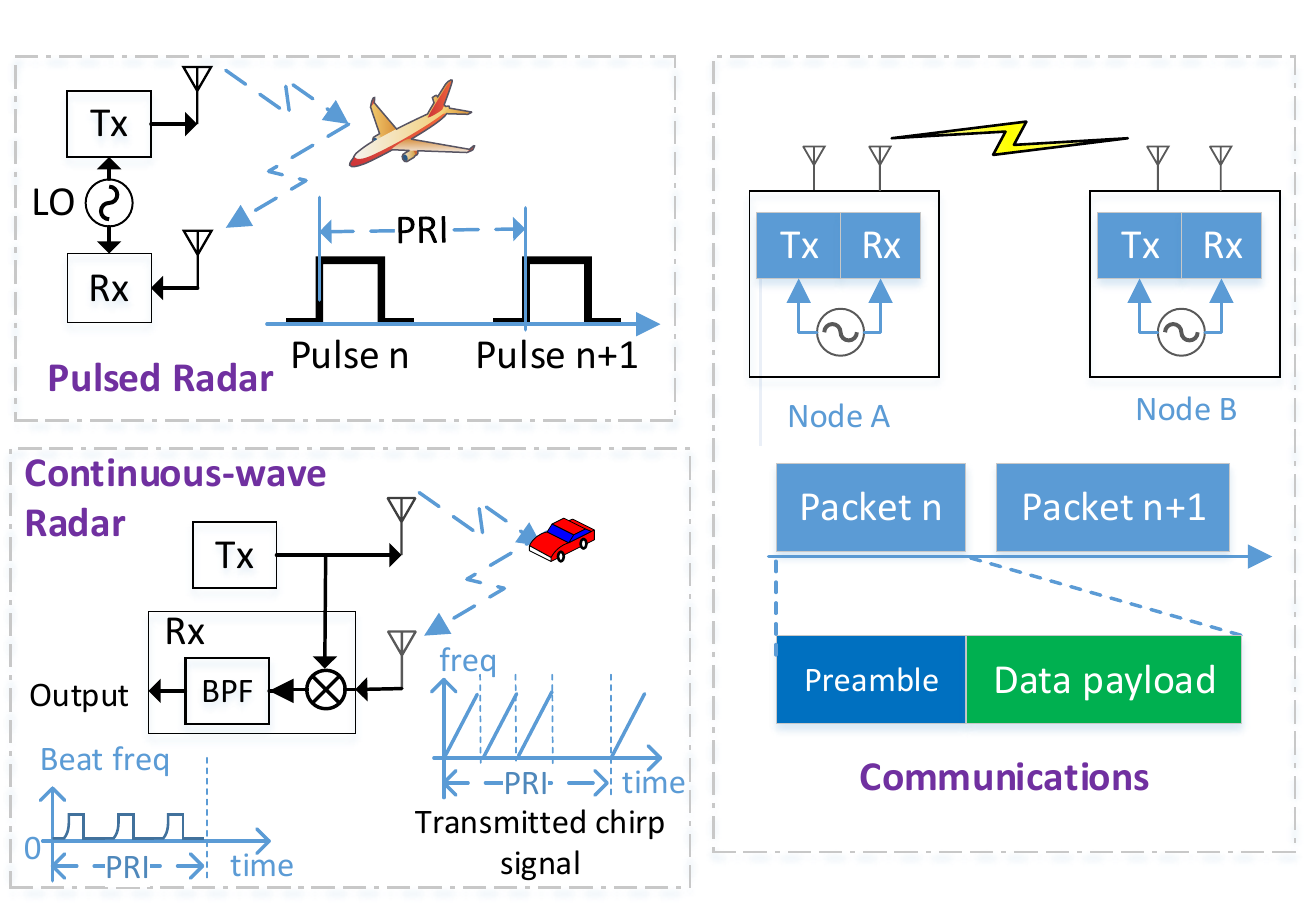}
	\caption{Illustration of basic pulsed radar, continuous-wave radar, and communication systems. Tx: transmitter; Rx: Receiver; BPF: bandpass filter; PRI: Pulse repetition interval.}
	\label{fig-basicsys}
\end{figure}

In this section, we describe some typical C\&R systems and their signal models. Since these systems and models are generally well known in respective areas, we only provide brief descriptions as necessary to  illustrate the JCR technologies. The basic pulsed radar, continuous-wave radar, and communication systems are illustrated in Fig. \ref{fig-basicsys}. We first present a beamspace channel model that can be used for both C\&R. We then describe signal models for C\&R separately, and highlight the differences between them. Signal models for JCR will be described in later sections, based on the models presented here. For the simplicity of notation, we assume the total transmitted power of all systems is $1$, unless noted otherwise. 
 
We consider a general system with $Q\geq 2$ nodes and each node has a uniform linear antenna array (ULA). Let $B$ denote the total signal bandwidth. For MIMO-OFDM systems, $B$ is divided into $N$ subcarriers and the subcarrier interval is $f_0=B/N$. The OFDM symbol period is $T_s=T_0+T_{cp}$ where $T_0=N/B$ and $T_{cp}$ is the period of cyclic prefix. 

In this section, for the clarity of illustration, we describe the signal models with reference to a general transmitter and receiver without indexing them. Later, when these models need to be differentiated for different nodes, we will add a subscript of the node index to the variables in the models.

\subsection{Beam-space Channel Models}\label{sec-channel}

Let the angle-of-departure (AoD) and angle-of-arrival (AoA) of a multipath be $\theta_{\ell}$ and $\phi_{\ell}$, $\ell\in[1,L]$, respectively. Assume a planar wave-front in signal propagation. The array steering/response vector of a ULA is given by
\begin{align*}
\av(M,\alpha)=[1,e^{j2\pi d/\lambda \sin(\alpha)},\cdots, e^{j (M-1)2\pi d/\lambda \sin(\alpha)}]^T,
\end{align*}
where $M$ is the number of antennas, $\lambda$ is the wavelength, $d$ is the interval of antennas, and $\alpha$ is either AoD or AoA.

For $M_{T}$ transmitting and $M_{R}$ receiving antennas, {the $M_{R}\times M_{T}$ time-domain baseband channel matrix at time $t$ can be represented as}
\begin{align}
{\Ht}(t)=\sum_{\ell=1}^L b_{\ell} \delta(t-\tau_{\ell}-\tau_o(t)) & e^{j2\pi (f_{D,\ell}+f_o(t)) t} \cdot \notag\\
&\av(M_{R},\phi_{\ell})\av^T(M_{T},\theta_{\ell}),
\label{eq-Ht1}
\end{align}
where for the $\ell$-th multipath, $b_{\ell}$ is its amplitude of complex value, accounting for both signal attenuation and initial phase difference; $\tau_{\ell}$ is the propagation delay; $f_{D,\ell}$ is the associated Doppler frequency; and $\tau_o(t)$  and $f_o(t)$ denote the potential time-varying timing offset and carrier frequency offset (CFO) due to possibly unlocked clocks between transmitter and receiver, respectively. Here, we ignore the ``beam squint'' effect in BF and assume $b_\ell$ is frequency independent. 

Equation (\ref{eq-Ht1}) represents a general channel model that can be used for both C\&R, although the physical meaning and names of these parameters are slightly different. Our description above is mainly based on the terminologies in communications. For radar sensing, $\{\tau_{\ell},f_{D,\ell},\phi_{\ell}, \theta_{\ell}, b_{\ell}\}$ are the \textit{sensing parameters} to be estimated. These parameters can be used to determine a target/reflector's spatial and moving information. In particular, $\phi_{\ell}$ and $\theta_{\ell}$ are the AoA and AoD of the target in relation to the receiving and transmitting ULA, resepctively; $\tau_{\ell}=R_\ell/c$ and $f_{D,\ell}=v_\ell f_c/c$ where $R_\ell$ is the signal propagation distance, $c$ is the speed of light, $v_\ell$ is the radial velocity of the reflector, and $f_c$ is the carrier frequency; and $b_{\ell}$ is related to the radar cross-section (RCS) and material property of the target. We define a \textit{coherent processing interval (CPI)} when all these parameters remain almost unchanged. The length of CPI depends on the mobility of objects in the channels and is typically a few milliseconds when objects move at speeds of tens of meters per second. 

For a broadband OFDM system, the frequency-domain channel matrix at the $n$-th subcarrier corresponding to  (\ref{eq-Ht1}) is 
\begin{align}
\Hf_{n}(t)=\sum_{\ell=1}^L b_\ell &e^{-j2\pi n(\tau_{\ell}+\tau_o(t))f_0}e^{j2\pi (f_{D,\ell}+f_o(t))t}\cdot \notag\\
&\av(M_{R},\phi_{\ell})\av^T(M_{T},\theta_{\ell}),
\label{eq-hfbasic}
\end{align}
where we have approximated the slightly varying phases due to Doppler frequency and CFO over one OFDM block as a constant value. For the $k$-th OFDM symbol with $t=kT_s$, we denote $\Hf_{n,k}=\Hf_n(t)_{t=kT_s}$. 

Note that for communications, we generally only need to know the composite values of the matrix ${\Ht}(t)$ or $\Hf_{n}(t)$. They can typically be obtained by directly estimating channel coefficients, or for OFDM, directly estimating at some subcarriers and obtaining the rest via interpolation. For radar sensing, however, the system needs to resolve the detailed channel structure and estimate the sensing parameters. Note that this beam-space channel model is also widely used in millimeter wave (mmWave) communication systems.

When the oscillator clocks of the transmitter and receiver are not locked/synchronized, both the timing offset $\tau_o(t)$ and CFO $f_o(t)$ are nonzero. The values of $\tau_o(t)$ can also be fast time-varying in terms of ranging due to crystal oscillator's instability, while CFO changes relatively slower. For example, for a typical clock stability of $20$ parts-per-million (PPM), the accumulated maximal variation of $\tau_o(t)$ over $1$ millisecond can be $20$ nanoseconds, which translates to a ranging error of $6$ meters. For communications, these offsets are generally not a big problem, as $\tau_o(t)$ can be absorbed into channel estimates after synchronization, and $f_o(t)$ can be estimated and compensated. Their residual values become relatively small compared to the baseband signal parameters. However, for radar sensing, when these offsets are unknown to the receiver, they can cause ambiguity in range and speed estimation, and become obstacles for processing signals across packets coherently. 

\subsection{Basic Communication Systems and Signals}\label{sec-comsys}

We consider a node transmitting $M_S$ spatial streams with $M_{T}\geq M_S$ antennas. The description here is with reference to a single user MIMO system, and it will be extended to multiuser MIMO later. 

\subsubsection{Single Carrier MIMO}

For a general single carrier (SC) MIMO system, we can represent the baseband signal vector at time $t$ from the transmitter as
\begin{align}
\xt(t)=\Pc \st(t),
\label{eq-xt}
\end{align}
where $\Pc$ is the spatial precoding matrix of size $M_{T}\times M_S$ and is typically fixed over a packet, and $\st(t)$ are the data vector of $M_S \times 1$. The symbols $\st(t)$ can be either the directly modulated constellation points that are unknown to the receiver, or known pilots. In the case of spread spectrum signals, each element in $\st(t)$ can be the product of a pseudo-random code and the constellation point (or pilot).

For a narrowband system, after propagating over the channel $\Ht(t)$, the received signals are given by
\begin{align} 
\yt(t)=\Ht(t)\circledast\xt(t)+\zt(t),
\label{eq-ytnb}
\end{align}
where $\circledast$ denotes convolution, and $\zt(t)$ is the AWGN.

\subsubsection{MIMO-OFDM}

For a MIMO-OFDM system, the baseband transmitting signals at subcarrier $n$ in the $k$-th OFDM symbol over all antennas can be represented as
\begin{align}
\xf_{n,k}=\Pc_{n,k} \st_{n,k},
\label{eq-mimoofdm}
\end{align}
where these variables are similarly defined as those in \eqref{eq-xt}, but could have different values for different subcarriers. In the case of MIMO-OFDMA, each node may be allocated to a resource block occupying groups of the antennas, subcarriers, and OFDM symbols, that are typically non-overlapping. These blocks can be discontinuous and irregular in these domains, which can cause significant challenges in sensing parameter estimation, as will be discussed in Section \ref{sec-sensing}. 

After propagating over the channel, the received frequency-domain baseband signals at subcarrier $n$ over all antennas are given by
\begin{align}
\yf_{n,k}=\Hf_{n,k}\xf_{n,k}+\zf_{n,k},
\end{align}
where $\zf_{n,k}$ is the AWGN.
 
\subsection{MIMO Radar Signals and Systems}\label{sec-mimoradar}

In a MIMO radar, the waveforms transmitted from different antennas are typically orthogonal, and the waveform can be either pulsed or continuous waveforms. The MIMO radar baseband waveform at a transmitter with $M_{T}$ antennas can be expressed as
\begin{align}
\mrt(t)= \sum_{m=1}^{M_{T}} \w_{m}\psi_{m}(t) = \W \bpsi(t),
\label{eq-mimo-rad}
\end{align}
where $\mrt(t)=\{x_{R,m}(t)\}, m=1,\ldots,M_T$ with $x_{R,m}(t)$ being the signal at the $m$-th antenna, $\psi_{m}(t)$ is the basic radar waveform, $\w_{m}$ is the $M_{T}\times 1$ precoding/BF vector, and $\W=[\w_{1},\cdots,\w_{M_{T}}]$ and $\bpsi(t)=\{\psi_{m}(t)\}$, $m=1, \cdots, M_{T}$. Note that in radar, the peak to average power ratio (PAPR) of the transmitted signal is typically required to be very low, so that high power efficiency can be achieved. Thus $\W$ is typically an identity matrix. 

The basic waveforms $\psi_{m}(t)$ can be any set of signals that are orthogonal in either one or multiple domains of time, frequency, space, and code. It can also take any waveform of pulsed and continuous-wave radars \cite{4383615,mimoradarbook}. In pulsed radar systems, short pulses of large bandwidth are transmitted either individually or in a group, followed by a silent period for receiving the echoes of the pulses, as can be seen from Fig. \ref{fig-basicsys}. Continuous-wave radars transmit waveforms continuously, typically scanning over a large range of frequencies. In both systems, the waveforms are typically non-modulated. 

Referring to the channel model in \eqref{eq-Ht1}, $\tau_o(t)$ and $f_o(t)$ are zeros for radar since the clocks for transmitter and receiver are typically locked. With the transmitting signal in \eqref{eq-mimo-rad}, the received noise-free radar signal is given by
\begin{align}
\yt_R(t)=\sum_{m=1}^{M_{T}}\sum_{\ell=1}^L& b_{\ell} \psi_{m}(t-\tau_{\ell})  e^{j2\pi f_{D,\ell} t} \cdot \notag\\
&\av(M_{R},\phi_{\ell})\av^T(M_{T},\theta_{\ell})\w_{m},
\end{align}
Applying matched filtering with $\psi_{m'}(t)$ to $\yt_R(t)$, we obtain
\begin{align*}
\rt_m(t)=\sum_{\ell=1}^L& b_{\ell} \rho_{m}(t-\tau_{\ell})  e^{j2\pi f_{D,\ell} t} 
\av(M_{R},\phi_{\ell})\av^T(M_{T},\theta_{\ell})\w_{m},
\end{align*}
where $\rho_m(t)$ is the non-zero output of the matched filtering of $\psi_{m}(t)$ when $m=m'$, as all other outputs for $m\neq m'$ are zeros. 

Assume that $\rho_m(t)=\rho(t)$ is the same for all $m\in[1,M_T]$. Staking all $\rt_m(t),m=1,\cdots, M_T$ to a matrix, we get
\begin{align*}
\Rt(t)=\sum_{\ell=1}^L& b_{\ell} \rho(t-\tau_{\ell})  e^{j2\pi f_{D,\ell} t} 
\av(M_{R},\phi_{\ell})\av^T(M_{T},\theta_{\ell})\W.
\end{align*}
When $\W$ is an identity matrix (if not, multiplying both sides with $\W^{-1}$), we can vectorize $\Rt(t)$ and get
\begin{align*}
\text{vec}(\Rt(t))=\sum_{\ell=1}^L& b_{\ell} \rho(t-\tau_{\ell})  e^{j2\pi f_{D,\ell} t} 
\av(M_{R},\phi_{\ell})\otimes\av^T(M_{T},\theta_\ell),
\end{align*}
where $\otimes$ denotes the Kronecker product.

In MIMO radars, particularly mono-static radars, the antenna intervals of transmitter and receiver, $d_T$ and $d_R$, are typically set as $d_T=M_{T}d_R$ or  $d_R=M_{T}d_T$. Then when $\phi_{\ell}=\theta_\ell$, which is generally true for a monostatic radar, we have $\av(M_{R},\phi_{\ell})\otimes\av^T(M_{T},\theta_\ell)=\av(M_RM_T, \phi_{\ell})$. This enables a MIMO radar to achieve the spatial resolution corresponding to a virtual ULA with $M_{T}M_{R}$ antennas \cite{mimoradarbook}. Note that the increased aperture of the virtual array is only meaningful when $\phi_{\ell}$ is related to $\theta_\ell$ in some way, although $\phi_{\ell}=\theta_\ell$ is not a necessary condition. 
 
In this paper, we mainly consider the following emerging MIMO radars, which have not been widely implemented in practice, but have great potentials for realizing JCR systems with balanced C\&R performance.

\subsubsection{MIMO-OFDM Radar}

In a MIMO-OFDM radar, the waveform $\psi_{m}(t)$ is in the form of time-domain OFDM signals. Without considering the cyclic prefix, the baseband signal of $\psi_{m}(t)$ can be represented as
\begin{align}
\psi_m(t)=\sum_{n\in\mathcal{S}_m} \tilde{w}_{m,n} e^{j2\pi nf_0t} g(t-kT_0),
\end{align}
where $\mathcal{S}_m$ is the set of used subcarriers, $g(t)$ is a windowing function, and $\tilde{w}_{m,n}$ can be a complex variable combining orthogonal coding, subcarrier-dependent spatial precoding/BF, and/or other processing such as PAPR reduction. 

To achieve orthogonality, $\mathcal{S}_m$s are typically selected to be orthogonal for different $m$. The set of subcarriers can be allocated in various forms, such as the interleaved pattern \cite{Lin15}, and nonequidistant subcarrier interleaving \cite{Hakobyan20}. Different allocations may lead to different signal properties and ambiguity functions. Alternatively, the orthogonality may also be achieved over time domain via the use of orthogonal codes, while different antennas share the subcarriers.

MIMO-OFDM signals for radar are very similar to those for communications, except that they are typically non-modulated and/or orthogonal across antennas. Hence MIMO-OFDM signals are excellent options for JCR systems. In particular, the training sequence in the preamble of MIMO-OFDM communication systems holds the desired characteristic of orthogonality, and can be directly used for radar sensing.

\subsubsection{Frequency-Hopping MIMO Radar and Frequency Agile Radar}\label{sec-farfh}
The frequency-hopping (FH) MIMO radar \cite{4602540} and the frequency agile radar (FAR) \cite{Alelsson07}, as well as its extensions to multicarrier \cite{Huang20apar}, all use the FH technologies - the total bandwidth is divided into many subbands, only a subset of subbands are used at a time, and the subset at each antenna randomly varies over time. FH can be implemented in either fast time or slow time, and in the form of either pulse or continuous-wave. We consider pulsed fast FH here, i.e., the signals are continuously transmitted with frequencies being changed rapidly and in multiple times over a PRI, followed by a silent period. Using FH leads to major advantages such as better security, and lower implementation cost by avoiding the use of costly instantaneous wideband components, while with negligible degradation in sensing performance, compared to using full bandwidth signals. FH-MIMO radar and FAR and its variants differ in how the frequencies are used at each antenna. Next, we briefly present their signal models, using notations similar to those for OFDM: $B$ for the radar bandwidth, $N$ for the number of sub-bands, and $T$ for the hop duration.

In a FAR, all antennas use one common frequency at a hop, and a BF weight is applied to each antenna so that the array forms steerable beam \cite{Alelsson07}. The basic concept of FAR is extended to multi-subband signaling in \cite{Huang20apar}, that is, a subset of more than one frequencies are used in each hop. In particular, in the multi-Carrier AgilE phaSed ArrayRadar (CAESAR) scheme proposed in \cite{Huang20apar}, the whole array is divided into multiple non-overlapped subarrays, and each antenna in one subarray only uses one common frequency from the frequency subset. CAESAR randomizes both the frequencies and their allocation among the antenna elements, and induces both frequency and spatial agility. It also maintains narrowband transmission from each antenna and introduces the BF capability. These capabilities make CAESAR more attractive than the original FAR. Let $\mathcal{A}_s$ denote the index set of the antennas in the $s$-th subarray, and $S$ be the number of total subarrays. Let $\mathcal{F}_k$ be the set of frequencies selected for hop $k$, and  $f_{k,m}\in\mathcal{F}_k$ denote the centroid frequency of the sub-band selected by antenna $m$ at hop $k$. The transmitted signal of CAESAR at the $m$-th antenna can be represented as
\begin{align}
\label{eq-caesar}
&\psi_{m}(t)=\tilde{w}_{k,m}e^{j2\pi f_{k,m}t}g(t-kT),\\
&\text{s.t.}\ \left\{\
\begin{array}{l}
f_{k,m}=f_{k,m'}, \text{when}\ \{m,m'\}\in \mathcal{A}_s, \forall s; \\
f_{k,m}\neq f_{k,m'}, \ \text{otherwise}.
\end{array}
\right.\notag
\end{align}
where $\tilde{w}_{k,m}$ is the BF weight.

Although FH-MIMO was developed before CAESAR, it can be treated as a special case of CAESAR where each subarray has only one antenna and each frequency is only used by one antenna, that is $S=M_T$ and $|\mathcal{A}_s|=1$. CAESAR can also be regarded as a generalization of FH-MIMO radar by introducing the BF capability. 

Both FH-MIMO radar and CAESAR are based on frequency division. They can also be realized on the framework of MIMO-OFDM, with the frequency hopping concept being introduced to subcarriers. 

\subsection{Major Differences between C\&R Systems and Signals}\label{sec-diffcs}

\begin{table*}[t]	
	\centering
	\caption{Brief comparison between C\&R signals and systems.}
	\begin{tabular}{|p{1.5cm}|p{8.2cm}|p{7cm}|}		
		\hline
		\textbf{Properties} & \textbf{Radar} & \textbf{Communications}  \\
		\hline
		\textbf{Typical Signal Waveforms}
		&
		
		Signals are unmodulated and have large bandwidth and Low PAPR; Orthogonal if MIMO-radar.
		&
		Mix of unmodulated (pilots) and modulated symbols; High PAPR; complicated and diverse signal waveforms.
		\\
		\hline
		\textbf{Signal Structure}
		&
		\begin{itemize}
			\item A silent period follows each pulse transmission in pulsed radar to allow the reception of echo signals;
			\item In continuous-wave radar, signals can be transmitted continuously, enabled by special hardware designs;
			\item Signal repeats every PRI within CPI to increase received signal power and enable Doppler frequency estimation.
		\end{itemize} 
		& 
		\begin{itemize}
			\item Typically packet-based. Typically no repetition;
			\item Packet length and interval can be time-varying;
			\item Signal may occupy discontinuous resources in time, frequency and space domains.	
		\end{itemize}
		
		\\
		\hline
		\textbf{Transmission Capability (Duplex)}
		&
		\begin{itemize}
			\item In continuous-wave radar, transmitted signal is used as local oscillator input at Rx to realize full duplex. This outputs ``beat'' signals only, characterizing the variation of the signals;
			\item Pulsed radar operates in half duplex mode with a silent period following each pulse transmission. 
		\end{itemize}		
		&
		Time division duplex or frequency division duplex mode.	Full duplex is immature for communications. Short-term solutions can be used to enable communication-centric JCR \cite{Zhangpmn20}. 
		
		\\
		\hline
		\textbf{Clock Synchronization}
		&
		Transmitter and receiver are clock-locked in most radar setups, including monostatic, bistatic and multi-static systems.
		& 
		Co-located transmitter and receiver share the same timing clock, but non-colocated nodes typically do not. 
		\\
		\hline
		\textbf{Receiver Signal Sampling}
		& A conventional continuous-wave radar samples received signals at a rate much smaller than the scanning bandwidth, proportional to the desired detection capability of the maximal ranging and moving speed. This makes information conveying difficult.
		& Sampling rate corresponds to the signal bandwidth. Full-bandwidth information available.
		
		\\
		\hline
		\textbf{Performance Metrics}
		& Detection probability, Cramer-Rao lower bound (CRLB), Mutual information (MI), Ambiguity function
		& Capacity, Rate, Spectral efficiency, Signal-to-interference-and-noise ratio (SINR), and Bit error rate (BER)
		\\
		\hline		
		
	\end{tabular}	
	\label{table:com}	
\end{table*}

As can be partially observed from the signal models in preceding subsections, there are some significant differences between C\&R systems and signals, despite of their potentials for integration. A brief comparison is summarized in Table~\ref{table:com}, where radar waveforms are mainly referred to traditional pulsed and continuous waveforms such as chirp. Next, we elaborate two aspects that have considerable impacts on the joint system design.

Firstly, it is a fundamental challenge to address the potential requirements for full duplex operation of JCR systems. On one hand, (mono-static) radar addresses the requirement for full duplex operation in mainly two approaches, as illustrated in Fig. \ref{fig-basicsys}, which may not be replicable in communication systems. One approach is typically applied in a pulsed radar, by applying a long silent period to receive echo signals, which essentially bypasses full-duplex operation and makes the radar work in a time-division duplex mode; the other is typically used in a continuous-wave radar, via using the transmitter signal as the local template signal to the oscillator at the receiver, and detecting only the ``beat'' signal, the difference between the transmitted and received signals. Such designs enable low-complexity and efficient radar sensing. However, they constrain the options of integrating communications and limit the achievable communication rates. For example, there is large uncertainty with the availability and bandwidth of the beat signal, hence information conveying will be unreliable. On the other hand, full-duplex operation is still immature for communications, and there is typically clock asynchronism between spatially separated transmitting and receiving nodes. These impose significant limits on integrating radar sensing into communications.

Secondly, C\&R signals are originally designed and optimized for different applications, and are generally not directly applicable to each other. Radar signals are typically designed to optimize localization and tracking accuracy, and enable simple sensing parameter estimation. The following properties of radar signals are desired: low PAPR to enable high-efficiency power amplifier and long-range operation; and an waveform ambiguity function with steep and narrow mainlobes for high resolution; Comparatively, communication signals are designed to maximize the information-carrying capabilities, and are typically modulated and packet-based. To support diverse devices and meet various quality-of-services requirements, communication signals can have complicated structures, with advanced modulations applied across time, frequency, and spatial domains, and being discontinuous and fragmented over these domains.  

These differences make the integration of C\&R an interesting and challenging task. As we will see from the following three sections, a good JCR design always seek and exploit the commonalities between C\&R, while taking these differences into consideration.  

\section{JCR: Communication-Centric Design}\label{sec-ccd}

In communication-centric JCR systems, radar sensing is integrated into existing communication systems as a secondary function. Revision and enhancement to communication infrastructure and systems may be required, but the primary communication signals and protocols largely remain unchanged.

Considering the topology of communication networks, communication-centric JCR systems can be classified into two types, namely, those realizing sensing in point-to-point communication systems and in large networks such as mobile networks. Two good examples are the IEEE 802.11ad JCR systems for vehicular networks \cite{Kumari20, Ali20, Duggal20} and the perceptive mobile networks \cite{8827589,Zhangpmn20}, respectively. They use the single carrier and multiuser-MIMO OFDM signals as described in Section \ref{sec-comsys}, respectively. Both are DFRC systems where a single transmitted signal is used for both C\&R.

In this section, we first describe the two types of DFRC systems, and then review some signal processing techniques for general communication-centric JCR systems.

\subsection{802.11ad DFRC Systems}\label{sec:802.11ad}

The 802.11ad standard defines a millimeter-wave packet communication system operating in the 60 GHz unlicensed band. There are three types of physical-layer (PHY) packets: single-carrier, OFDM, and control, with OFDM being optional. The preamble in each packet is the main signal that has been exploited for radar sensing \cite{Kumari17,Grossi18,Kumari20, Ali20, Duggal20} in an 802.11ad DFRC system.  Although the DFRC system can be applied in many scenarios, it has been mainly investigated for vehicular networks. In a typical setup, the sensing receiver is co-located with the DFRC transmitter, using two separated analog arrays. The DFRC device can be located either on the road side unit (RSU) or on a vehicle.

Referring to \eqref{eq-Ht1}, the noise-free time-domain echo signal at the sensing receiver can be represented as
\begin{align}
y(t)=\sum_{\ell=1}^L h_\ell(t) s(t-\tau_{\ell}) e^{j2\pi f_{D,\ell} t},
\label{eq-11ad}
\end{align}
where $h_\ell(t)\triangleq b_\ell\w_R(t)^T\av(M_R, \theta_\ell)\av^T(M_T,\theta_\ell)\w_T(t)$, $\w_T(t)$ and $\w_R(t)$ are the beamforming vector in the transmitter and receiver respectively, and the AoA and AoD are assumed to be the same. Note that the clock between the transmitter and sensing receiver is locked, therefore $\tau_o(t)=0$ and $f_o(t)=0$ in~\eqref{eq-Ht1}. The primary goal of sensing here is estimating the location and velocity of objects via estimating $\tau _\ell$, $\theta_\ell$, and $\f_{D,\ell}$.

The three PHYs in 802.11ad have a similar preamble structure, consisting of short training field (STF) and channel estimation field (CEF). The STF consists of tens of repeated 128-sample Golay sequences, followed by its binary complement. The CEF consists of two 512-sample Golay complementary pair, which has the property of perfect aperiodic autocorrelation, i.e., $s(t-\tau)\circledast s(t)\neq 0$ if and only if $\tau=0$. Both the STF and CEF can be used for sensing, in either a hierarchical or joint manner \cite{Kumari17}. The hierarchical strategy processes the STF and CEF separately, exploiting their respective properties. For example, the repetition pattern of STF is typically used for packet detection in communications, and hence it is ideal for target detection in sensing; while the perfect aperiodic autocorrelation of CEF can lead to excellent channel estimation and sensing performance, based on, e.g., the generalized likelihood ratio test \cite{Grossi18}. The joint strategy uses both STF and CEF for common tasks of sensing, based on, e.g., matched filtering \cite{Kumari17}. The sensing performance bounds are also derived in \cite{Kumari17,Grossi18}. More advanced sensing algorithms will be discussed in Section \ref{sec-sensing}.

Both the single-carrier PHY, which has an identical preamble with the OFDM PHY, and the control PHY have been explored for sensing \cite{Grossi18,Kumari17}. There are some differences between their sensing efficiency. On one hand, in the standard, a beamforming training protocol is defined to align the transmit and receive beams, using beam scanning and the control-PHY signals. Single-carrier or OFDM PHY signals are typically used after beamforming training. On the other hand, the control PHY has a longer STF. So in terms of sensing, the control PHY enables a wider field-of-view (FoV) and potentially better accuracy, while the other two are generally limited to the fixed direction of communications. A multibeam approach, as will be discussed in Section \ref{sec-multibeam}, can be applied to relax this limitation.

\subsection{Mobile network DFRC Systems}\label{sec-pmn}
\begin{figure}[t]
	\centering
	\includegraphics[width=1\linewidth]{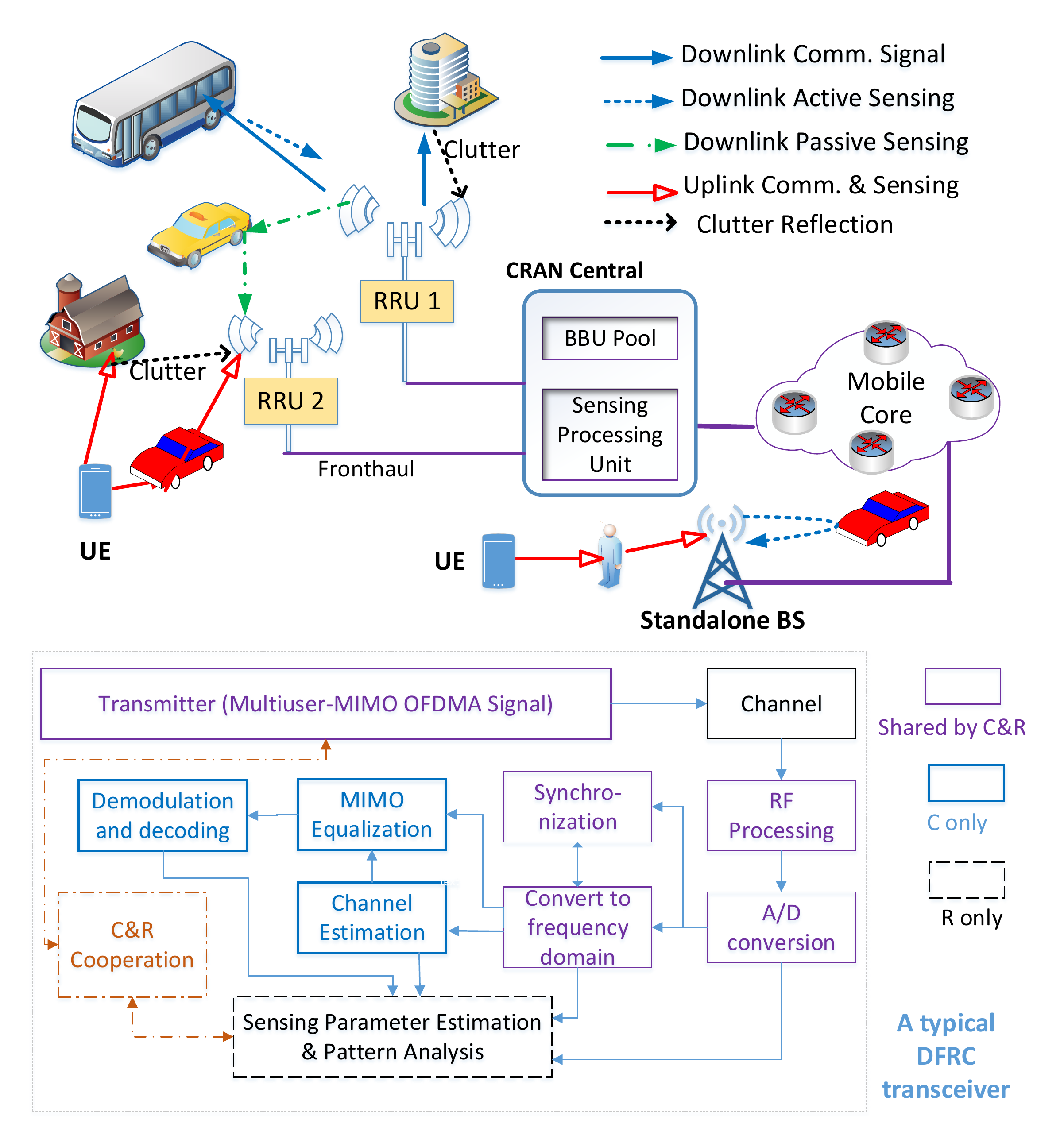}
	\caption{Network setup of PMNs (top subfigure) and the brief system block diagram of a DFRC transceiver based on MIMO-OFDM (bottom subfigure).}
	\label{fig-PMN}
\end{figure}

In \cite{8827589}, the framework of perceptive mobile networks (PMNs) is introduced by applying the JCR, more specifically DFRC, techniques, to cellular networks. Downlink sensing and uplink sensing are defined, corresponding to sensing using the received downlink and uplink communication signals, respectively, as illustrated in Fig. \ref{fig-PMN}. In the scenario of cloud radio access networks (CRANs) where distributed remote radio units (RRUs) cooperatively communicate with user equipment (UE), the received downlink communication signals from one RRU itself and other cooperative RRUs can be used for downlink active sensing and downlink passive sensing, respectively. 

Extend the single user MIMO-OFDM model in Section \ref{sec-comsys} to multiuser MIMO-OFDMA. Suppose that one node receives signals transmitted from a set of nodes $q, q\in \mathcal{Q}_T$, and uses the signals for sensing. Let $Q_T$ be the cardinality of $\mathcal{Q}_T$. Referring to the transmitting signal model in \eqref{eq-mimoofdm} and the channel model in \eqref{eq-hfbasic}, we can represent the received noise-free $k$-th OFDM symbol at the $n$-th subcarrier as 
\begin{align}
\yf_{n,k}
=\sum_{q\in \mathcal{Q}_T}\sum_{\ell=1}^{L_q}& b_{q,\ell} e^{-j2\pi n (\tau_{q,\ell}+\tau_{o,q,k})f_0}e^{j2\pi k (f_{D,q,\ell}+f_{o,q,k}) T_s}\cdot\notag\\
& \quad\av(M_R,\phi_{q,\ell})\av^T(M_q,\theta_{q,\ell})\xf_{q,n,k}  
\label{eq-yf1}
\end{align}   
Scenarios represented by this model are exemplified below:
\begin{enumerate}
	\item \textit{Downlink sensing in a standalone base station (BS)}: This is the case where the BS uses its own reflected transmitted signals for sensing, similar to a mono-static radar. In this case, $Q_T=1$ and $\tau_{o,q,k}=f_{o,q,k}=0$;
\item \textit{Uplink sensing in a standalone BS}:  $\mathcal{Q}_T$ denotes the set of $Q_T$ UEs sharing the same subcarriers via SDMA, and $\tau_{o,q,k}\neq0$ and $f_{o,q,k}\neq 0$.. Each UE only occupies partial of the total subcarriers;
\item \textit{Downlink sensing in an RRU}:  $\mathcal{Q}_T$ denotes the set of RRUs whose downlink communication signals are seen by the sensing RRU, including its own echo signals.
\end{enumerate}
The sensing can be based on \eqref{eq-yf1} with the signals $\xf_{q,n,k}$ corresponding to both the pilots and data symbols, as will be detailed next. For more details on the signals usable for sensing, as well as solutions to the full-duplex problems in PMNs, the readers are referred to \cite{8827589,Zhangpmn20}.
\subsection{Sensing Parameter Estimation} \label{sec-sensing}

Sensing parameter estimation in communication-centric JCR is generally different to that in traditional radar systems, due to the significant differences between the two types of signals as described in Section \ref{sec-system}. Next, referring to the MIMO-OFDM signal models in Section \ref{sec-pmn}, we review key techniques in sensing parameter estimation. Most of them are also applicable to single carrier systems such as the IEEE802.11ad DFRC. We first discuss two important problems to be resolved, before we review optional sensing algorithms.

\subsubsection{Direct and Indirect Sensing}

The first problem is how to deal with $\xf_{q,n,k}$ in the received signals, which can represent either known pilots or (unknown) data payload in a packet. Both can be used for sensing. For unknown data payload, it can be demodulated after channel estimation, as in conventional communications. Using data payload can significantly extend the sensing capability such as the range, as it is much longer than the pilot. Here, the receiver is assumed to know $\xf_{q,n,k}$ or its estimate via demodulating $\sfd_{q,n,k}$. For multiuser-MIMO signals, for example, signals received at an RRU from multiple RRUs in downlink sensing, or signals received at a standalone BS from multiple UEs, we can use two methods to formulate the estimation problem. 

One method, which may be called as \textit{direct sensing}, directly feeds the received signals to sensing algorithms. In some cases, e.g., sensing using the data payload in a MIMO system, this is the only option as $\xf_{q,n,k}$ cannot be readily removed even when they are known. The presence of $\xf_{q,n,k}$ often limits the optional algorithms for sensing parameter estimation.

Let us have a look at one example in \cite{8827589}, where direct sensing is conducted via the block compressive sensing (CS) techniques \cite{6415293}, and the symbols $\xf_{q,n,k}$ are used as part of the sensing matrix. For the clarify of presentation, we consider the case of $Q_k=1$ in \eqref{eq-yf1}, and ignore the timing offset $\tau_{o,q,k}$. Ignore the noise and rewrite \eqref{eq-yf1} in a more compact matrix form as
\begin{align}
\yf_{n,k}=\A(M_R,\bm{\phi}_q)\C_n\D_{k}\A^T(M_q,\bm{\theta}_q)\xf_{q,n,k},
\label{eq-active}
\end{align}   
where
$\A(M_R,\bm{\phi}_q)$ and $\A(M_q,\bm{\theta}_q)$ are matrices with the $\ell$-th column being $\av(M_R, \phi_{q,\ell})$ and $\av(M_q,\theta_{q,\ell})$), respectively, and $\D_k$ and $\C_n$ are diagonal matrices with the diagonal element being $b_{q,\ell} e^{j2\pi kf_{D,q,\ell} T_s}$ and $ e^{-j2\pi n \tau_{q,\ell}f_0}$, respectively. 

In order to apply sensing algorithms, we need to re-organize signals so that we can stack more measurements over the same domain. Consider the case of collecting samples from all the subcarriers for the estimation of delay and AoA. Take the transpose of $\yf_{n,k}$ in (\ref{eq-active}), and rewrite it as
\begin{align}
\yf_{n,k}^T 
&=\xf^T_{q,n,k}(\cv^T_n\otimes\I_{M_q}) \V_q\A^T(M_R,\bm{\phi}_q).
\label{eq-yfdelay}
\end{align} 
where $\cv_n$ is a column vector containing the diagonal elements of $\C_n$, $\I_{M_q}$ is an $M_q\times M_q$ identity matrix, and $\V_q$ is a block diagonal matrix
\begin{align*}
\V_q=\text{diag}\{b_\ell e^{-j2\pi k f_{D,q,\ell}T_s}\av(M_q,\theta_{q,\ell})\}, \ell=1, \ldots, L_q.
\end{align*} 

We have now separated signals $\xf^T_{q,n,k}(\cv^T_n\otimes\I_{M_q})$ that depend on $n$ from those on other variables. We can then stack the row vectors $\yf_{n,k}^T$ from all available subcarriers to a matrix, and obtain
\begin{align}
\Yf_{k}&\triangleq[\yf_{1,k},\cdots,\yf_{n,k},\cdots]^T 
= \W\V_q\A^T(M_R,\bm{\phi}_q),
\label{eq-vat}
\end{align}
where the $n$-th row of $\W$ is $\xf^T_{q,n,k}(\cv^T_n\otimes\I_{M_q})$.

The signal model in \eqref{eq-vat} enables the applications of both 1-D multi-measurement vector (MMV) CS and 2-D CS techniques. For 1-D MMV CS, $\cv_n$ is expanded to a quantized on-grid vector, and then $\W$ is used as the sensing matrix. The delay and $\V_q\A^T(M_R,\bm{\phi}_q)$ will then be the outputs of the algorithm, and the Doppler frequency and AoA can be further estimated from the estimates of $\V_q\A^T(M_R,\bm{\phi}_q)$. For 2D CS, both delay and AoD can be estimated together by expanding both $\cv_n$ and $\A(M_R,\bm{\phi}_q)$ to on-grid models. It is easy to see that if we swap the terms of Doppler frequency and delay in \eqref{eq-yfdelay},  samples across OFDM symbols can be stacked and the Doppler frequencies can be estimated first.

The direct sensing method has a high computational complexity. Due to the presence of $\xf_{q,n,k}$, the applicable sensing solutions are also limited. However, it is the only option, when $\xf_{q,n,k}$ cannot be removed due to, e.g, insufficient measurements. 
 
The other method, \textit{indirect sensing}, first estimate the elements of channel matrix for each node. It decorrelates signals from multiple nodes, removes $\xf_{q,n,k}$ or $\sfd_{q,n,k}$ from the received signals, and applies sensing parameter estimation to the estimated channel matrix. In multi-user MIMO systems, referring to \eqref{eq-yf1}, this can be achieved by decorrelating signals collected from $K$ $\yf_{n,k}$s, $k=k',k'+1,\cdots,k'+K-1, K\geq M_TQ_T$ at subcarrier $n$. Mathematically, this can be represented as
\begin{align}
&\left[\Hf_{1,n,k}, \cdots, \Hf_{Q_T,n,k}\right]\notag\\
=&\left[\breve{\xt}_{n,1},\cdots,\breve{\xt}_{n,K}\right]^{-1}\left[\yf_{n,1},\cdots,\yf_{n,K}\right],
\label{eq-hfcol}
\end{align}
where $\breve{\xt}_{n,k}=[\xf_{1,n,k}^T,\cdots,\xf_{Q_T,n,k}^T]^T$ is a $M_TQ_T\times 1$ vector. Note that the channel matrix is assumed to be constant over this interval. Equation \eqref{eq-hfcol} indicates that the decorrelation is only possible when $K\geq M_TQ_T$ $\breve{\xt}_{n,k}$s are available during a CPI and the inversion of $\left[\breve{\xt}_{n,1},\cdots,\breve{\xt}_{n,K}\right]$ exists. 

The decorrelation involves high computational complexity due to matrix inversion and may cause significant noise enhancement, unless $\xf_{q,n,k}$ or $\sfd_{q,n,k}$ are orthogonal. Hence indirect sensing is particularly suitable for training and pilot symbols which are typically orthogonal. After $\xf_{q,n,k}$ is removed, we can equivalently work on the single user channel matrix $\Hf_n(t)$ in \eqref{eq-hfbasic}. This can largely simplify sensing parameter estimation, and offer great flexibility in problem formulation. Note that if the precoding matrix $\Pc_{n,k}$ is unknown to the receiver, $\av^T(M_{T},\theta_{\ell})$ in \eqref{eq-hfbasic} will be replaced by $\av^T(M_{T},\theta_{\ell})\Pc_{n,k}$. This will make the estimation of AoD challenging. 

\subsubsection{Clutter Removal} 

The second problem is how to deal with clutter signals that are useless for sensing. Communication systems are typically deployed in an environment with dense multipath, where many signal propagation paths involve static objects only and are not of the interest of sensing. Removing clutter before sensing may distort the signal, however, it can largely reduce the number of parameters to be estimated. In this sense, it is generally a better strategy to remove clutter before the application of sensing algorithms, using, e.g., background subtraction or filtering techniques. For detailed discussions on potential clutter removal techniques, the readers are referred to \cite{8827589}.

\subsubsection{Sensing Algorithms}

We now discuss options for sensing algorithms based on the indirect method. Traditional radar typically applies matched filtering for sensing parameter estimation, which has also been adopted in some DFRC systems, e.g., 802.11ad DFRC \cite{Kumari17,Kumari20, Duggal20}. However, the accuracy and resolution capability of these methods largely depend on the signal correlation properties (i.e., ambiguity functions). More options that are less affected by the correlation property can be explored for communication-centric DFRC signals.

From the decorrelated estimates of  $\Hf_{n,k}$s in \eqref{eq-hfbasic}, we can represent the $(m_R,m_T)$-th element in $\Hf_{n,k}$ as
\begin{align}
\hfi_{n,k,m_R,m_T}&=\sum_{\ell=1}^L b_\ell e^{-j2\pi n\tau_{\ell}f_0}e^{j2\pi k f_{D,\ell}T_s}\notag\\
&\quad\cdot e^{j2\pi d_Rm_R\sin(\phi_\ell)/\lambda}e^{j2\pi d_Tm_T\sin(\theta_\ell)/\lambda},
\label{eq-hr}
\end{align}
where $m_R$ and $m_T$ represent the indexes of the receiving and transmitting antennas, respectively. This is also known as a 4-D \textit{Harmonic retrieval} problem \cite{Nion10}, where the observation signals in each domain can be represented as a Vandermonde matrix when the samples are equally spaced. The 4-D harmonic retrieval problem can be reduced to multiple-snapshot lower-dimensional problems by combining one or more of the exponential functions with the unknown variable $b_\ell$. Thus, we can rewrite them to different matrix and Tensor forms so that sensing parameters can be estimated in different ways and orders. 

Since solutions to the classical harmonic retrieval problems are generally well studied, we ignore the details and only provide a comparison of typical techniques  in Table \ref{table:method} with reference to our sensing problems. More details of such techniques for JCR can be referred to \cite{ Nion10, 7833233, 8827589,8905229,Liu20super}. 

When selecting sensing algorithms, the following issues need to be further considered:
\begin{enumerate}
	\item Modern communication signals are typically very complicated in terms of resource usage, and they may be discontinuous in one and more domains of space, frequency, and time. See \cite{Zhangpmn20} for the available sensing signals and their properties in 5G-based PMNs. This requires sensing algorithms with the capability of processing discontinuous and varying-interval samples;
	\item Higher-dimension algorithms can generally identify more parameters and achieve better estimation performance at the cost of higher complexity. However, when there are insufficient samples in one domain, it could be better to use lower-dimension algorithms.
\end{enumerate} 

\begin{table*}[t]
	
	\centering
	\caption{Comparison of Sensing Parameter Estimation Algorithms}
	\begin{tabular} {|m{2.5cm}|m{7.1cm}|m{7.1cm}|}
		
		\hline
		\textbf{Algorithms} & \textbf{Advantages} &\textbf{Main limitations} \\
		\hline
		
		Periodogram such as 2D DFT (typically based on the outputs of matched filtering)
		&
		Traditional technique. Simple and easy to implement. May be used as the starting point for other algorithms.
		&
		\begin{itemize}
			\item Low resolution;
			\item Generally require a full set of continuous samples in each domain, which may not always be satisfied.
		\end{itemize}	
		\\
		\hline
		
		Maximal Likelihood Estimation \cite{Grossi18} 
		&
		Statistically optimal formulation; Particularly suitable for low-dimension signals.
		& 
		Typically require searching to find the solutions and hence complexity is high; Complexity also increases with signal dimensions exponentially. 
		\\
		\hline
		Subspace methods such as ESPRIT and MUSIC \cite{Liu20super,ni2020uplink}
		&
		\begin{itemize}
			\item Separate signal and noise subspaces and hence is resilient to noise;
			\item ESPRIT can achieve very high resolution and can do off-grid estimation;
			\item MUSIC can flexibly work with non-continuous samples.
		\end{itemize}
		
		&
		\begin{itemize}
			\item ESPRIT requires a large segment of consecutive samples, which may not always be satisfied;
			\item Resolution of MUSIC depends on searching granularity;
			\item High complexity associated with singular value decomposition.			
			\end{itemize}
		\\
		\hline
		Compressive sensing (On-grid) \cite{7833233,8827589}
		&
		\begin{itemize}
			\item Flexible and does not require consecutive samples;
			\item Various recovery algorithms available, allowing good tradeoff between complexity and performance;
			\item Different dimensions of formulation can be used, adapting to sensing requirements and conditions;
			\item Dense dictionaries can be used to improve resolution. 
		\end{itemize}
		 		
		&
		Although it may even work well for estimating a small amount of off-grid parameters, performance can degrade significantly when the number of parameters to be estimated is large.
		
		\\
		\hline
		Compressive Sensing (Off-grid) such as atomic norm minimization \cite{7833233}
		&
		Have all the advantages of on-grid CS algorithms. Capable of estimating off-grid values.
		
		&
		Limitation in real time operation due to very high complexity. Still require sufficient separation between parameter values.
		\\
		\hline
		Tensor based algorithms \cite{Nion10}
		&
		High-dimension formulation and estimation are made easy.
		Reduce computational complexity and provide capability
		in resolving multipath with repeated parameter values.
		
		&
		Need to be combined with other algorithms such as ESPRIT and CS, thus facing their inherent problems.
		\\
		
		\hline
		
	\end{tabular}
	
	\label{table:method}
	
\end{table*}

\subsection{Resolution of Clock Asynchrony}\label{sec-resamb}

As described in Section \ref{sec-channel}, when the oscillator clocks between the transmitter and the sensing receiver are not locked, the timing offset $\tau_o(t)$ and CFO $f_o(t)$ in \eqref{eq-Ht1} and \eqref{eq-hfbasic}, are typically non-zero and time-varying. They can directly cause ambiguity in range and speed estimation. They also prevent from aggregating measurements over a relatively long interval, e.g., preamble signals from two packets, for joint processing, which is otherwise important for parameter estimation, particularly Doppler frequencies. This is a critical and challenging problem in communication-centric JCR systems. 

There have been a limited number of works that address this problem in passive WiFi sensing \cite{passive10, IndoTrack, widar2.0}, based on the cross-antenna cross-correlation (CACC) method. The basic assumption is that timing offsets and CFO across multiple antennas in the receiver are the same, as the same clock is used. Therefore these offsets can be removed by computing the cross-correlation between signals from multiple receiving antennas. Considering a single transmitter with a single antenna and referring to \eqref{eq-yf1}, the received noise-free signal at the $m$-th antenna can be rewritten as
\begin{align*}
\tilde{y}_{n,k,m}=\sum_{\ell=1}^{L} b_{\ell} e^{-j2\pi n (\tau_{\ell}+\tau_{o,k})f_0}&e^{j2\pi k (f_{D,\ell}+f_{o,k}) T_s} \dot \notag\\
&e^{jm\pi \sin(\phi_\ell)}\tilde{x}_{n,k}.
\end{align*}
Let the $m_0$-th antenna be the reference. Computing the cross-correlation between $\tilde{y}_{n,k,m}$ and $\tilde{y}_{n,k,m_0}$ yields
\begin{align}
\label{eq-crsscr}
&R(n,k,m)=\tilde{y}_{n,k,m}\tilde{y}_{n,k,m_0}^*\\
=&\sum_{\ell_m=1}^{L} \sum_{\ell_{m_0}=1}^{L} b_{\ell_m}b_{\ell_{m_0}}^* e^{-j2\pi n (\tau_{\ell_m}-\tau_{\ell_{m_0}})f_0} \notag\\
&\cdot e^{j2\pi k (f_{D,\ell_m}-f_{D,\ell_{m_0}}) T_s} e^{jm\pi (\sin(\phi_{\ell_m})-\sin(\phi_{\ell_{m_0}}))}|\tilde{x}_{n,k}|^2\notag.
\end{align}
Note that in \eqref{eq-crsscr}, $\tau_{o,k}$ and $f_{o,k}$ are removed. However, cross-correlation causes doubled terms and sensing parameters become relative. 

Two assumptions, which limit the applications, are necessary for subsequent processing: (1) The transmitter and sensing receiver are relatively static and the relative location of the transmitter is known to the receiver; and (2) there exists a line-of-sight (LOS) path between them and it has much larger magnitude than non-LOS paths. The cross-product between LOS paths is invariant over the CPI and can be removed by passing $R(n,k,m)$ through a high pass filter. The cross-terms between NLOS and LOS paths thus dominate in the output of the filter. The sensing parameters can then be estimated, with respect to the known parameters of the LOS path.  

However, the outputs after CACC contain cross-product terms that include signals and their images, and hence the number of unknown parameters to be estimated is actually doubled. This will not only cause degraded estimation accuracy, but also ambiguity between the actual value and its image. The authors in \cite{IndoTrack} proposed an \textit{add-minus} method to suppress the image signals by adding a constant to $\tilde{y}_{n,k,m}$ and subtracting another one to $\tilde{y}_{n,k,m_0}$. However, this method is found to be susceptible to the number and power distribution of static and dynamic signal propagation paths. To resolve these problems, a method is proposed in \cite{ni2020uplink} by introducing a \textit{mirrored MUSIC} algorithm.

Observing from \eqref{eq-crsscr}, we can see that the relative delays and Doppler frequencies have values symmetric to zero. Exploiting this symmetric, the mirrored MUSIC algorithm first constructs new signals from the CACC outputs, or the outputs of a further high pass filter to remove the dominating static components, and then define new basis vectors for MUSIC algorithms. Both of the new signals and basis vectors are constructed by adding the original ones with their sample-reversed versions. The mirrored MUSIC algorithm equivalently reduces the number of unknown parameters by half, and is shown to significantly improve the estimation accuracy, and simplify the ambiguity resolution problem associated with image signals.
	
\subsection{Sensing Assisted Communications}
In many location-aware services and applications, e.g., V2X network, sensing and communication are recognized as a pair of intertwined functions, where communication-centric JCR can be applied so as to reduce the costs and improve spectral-, energy-, and hardware-efficiency. In addition to those general advantages, one may also leverage the sensing results to facilitate communication, where significant performance gain can be obtained over that of the conventional communication-only schemes. Examples are sensing assisted secure communications \cite{Su20Secure} and BF \cite{859,9162000,Liu20,Yuan20Bayesian}. Below we briefly review the recent state-of-the-art on sensing-assisted BF.

BF for mmWave communications relies on the AoA and AoD parameters of signal propagation paths. With such angular information at hand, the Tx and the Rx can accurately align their beams, such that a high-quality communication link can be established. Conventionally, the AoA and AoD are acquired by {\emph {beam training}} or {\emph {beam tracking}} techniques, i.e., communication-only approaches. The basic idea of these schemes is to send pilots from the Tx to the Rx before data transmission, with each pilot being beamformed to a different direction. The Rx then estimates the angles from the received pilot signal and feeds them back to the Tx. 

It can be observed from the above procedure that there exists an inherent tradeoff between the estimation accuracy and the signalling overhead. Indeed, the beam information can be more precisely attained by transmitting more pilots, which, however, is at the price of reducing the effective transmission time of the useful data, as well as of increasing the latency. This problem is particularly pronounced in high-mobility V2X scenarios, where the trained beams can be easily outdated. To cope with this issue, radar sensing is exploited to enhance the BF performance via providing accurate positioning information for vehicles, such that the beam search interval can be narrowed down. Recent results show that with the aid of an extra radar sensor mounted on the road side unit (RSU), the beam training overhead incurred in the vehicle-to-infrastructure (V2I) links can be significantly reduced to 6.5\% of that of the Global Navigation Satellite System assisted approach \cite{859,9162000}.  

As a step beyond using stand-alone radar sensors, JCR signalling is envisioned to play a unique role in V2I communications, offering not only considerable reduction in the overhead, but also the capability to beamform towards predicted directions of the vehicles, in order to adapt the fast-changing vehicular channels. To show this, let us consider a mmWave RSU equipped with $M_T$ transmit and $M_R$ receive antennas, which acts as a mono-static radar, and is serving a single-antenna vehicle driving at a nearly constant speed on a straight road. We assume that the RSU communicates with the vehicle over a single LoS path, and that all the antenna arrays are adjusted to be parallel to the road. As a consequence, the AoA equals to the AoD in the V2I LoS channel.

At the $k$th epoch, the RSU transmits a DFRC signal $\mathbf{x}_k\left(t\right) = {{\mathbf{f}}_k}s_k\left( t \right)$ from the RSU to the vehicle, with ${{\mathbf{f}}_k}$ being the JCR beamformer, and $s_k\left( t \right)$ being the data stream. The signal is partially received by the vehicle's antenna array, and is partially reflected back to the RSU. The received echo signal can be expressed in the form of 
\begin{equation}\label{measurement1}
\begin{gathered}
  {{\mathbf{y}}_R}\left( t \right) = b_k{e^{j2\pi {f_{D,k}} t}}{\mathbf{a}}\left( {{M_R},{\theta _k}} \right){{\mathbf{a}}^T}\left( {{M_T},{\theta _k}} \right) \cdot  \hfill \\
  \quad\;\;\;\;\;\;\;\;\;\;\;\;\;\;\;{{\mathbf{f}}_k}s_k\left( {t - {\tau _k}} \right) + {\mathbf{z}}_R\left( t \right), \hfill \\ 
\end{gathered}
\end{equation}
where the beamformer $\mathbf{f}_k$ is designed based on a predicted angle, which is $
{{\mathbf{f}}_k} = {{\mathbf{a}}^ * }\left( {{M_T},{{\hat \theta }_{k\left| {k - 1} \right.}}} \right)$.
Moreover, ${{\hat \theta }_{k\left| {k - 1} \right.}}$ is the $k$th predicted angle based on the $\left(k-1\right)$th estimate, and $b_k$, $f_{D,k}$, $\theta_k$, and $\tau_k$ denote the reflection coefficient, the Doppler frequency, the AoA, and the round-trip delay for the vehicle at the $k$th epoch, respectively. In particular, $f_{D,k}$ and $\tau_k$ can be further written as functions of the distance $d_k$, the velocity $v_k$, and the AoA $\theta_k$, which are ${f_{D,k}} = \frac{{2{v_k}\cos {\theta _k}{f_c}}}{c},{\tau _k} = \frac{{2{d_k}}}{c}$. By matched-filtering (\ref{measurement1}) with a delayed and Doppler shifted counterpart of $s_k\left(t\right)$, one obtains the estimates of the Doppler and the time-delay, denoted as $\hat{f}_{D,k}$ and $\hat{\tau}_k$. Let us denote the vehicle's state as ${\mathbf{q}_k} = {\left[ {{\theta _k},{d_k},{v_k},{b_k}} \right]^T}$. The sensing measurement and the vehicle state can be connected by a function ${{{\mathbf{\tilde r}}}_k} = {\mathbf{h}}\left( {{{\mathbf{q}}_k}} \right) + {{\mathbf{z}}_k}$, which can be expanded as \cite{Liu20}
\begin{equation}\label{measurement2}
\begin{gathered}
  \left\{ \begin{gathered}
  {{{\mathbf{\tilde y}}}_R} = {b_k}{E_k}{\mathbf{a}}\left( {{M_R},{\theta _k}} \right){{\mathbf{a}}^T}\left( {{M_T},{\theta _k}} \right){{\mathbf{a}}^*}\left( {{M_T},{{\hat \theta }_{k\left| {k - 1} \right.}}} \right) + {{\mathbf{z}}_\theta }, \hfill \\
  {{\hat f}_{D,k}} = \frac{{2{v_k}\cos {\theta _k}{f_c}}}{c} + {z_f}, \hfill \\
  {{\hat \tau }_k} = \frac{{2{d_k}}}{c} + {z_\tau }, \hfill \\ 
\end{gathered}  \right. \hfill \\ 
\end{gathered}
\end{equation}
where $E_k$ is the matched filtering gain, and $\mathbf{z}_k = \left[z_\theta^T,z_f,z_\tau\right]^T$ denotes the measurement noise.

Recalling the assumption that the vehicle is driving on a straight road with a nearly constant speed, the state transition can be modeled as ${{\mathbf{q}}_k} = {\mathbf{g}}\left( {{{\mathbf{q}}_{k - 1}}} \right) + {{\bm{\omega }}_k}$, which can be expressed in the form of \cite{Liu20}
\begin{equation} \label{state_model}
\begin{gathered}
  \left\{ \begin{gathered}
  {\theta _k} = {\theta _{k - 1}} + d_{k - 1}^{ - 1}{v_{k - 1}}\Delta T\sin {\theta _{k - 1}} + {\omega _{\theta ,k}}, \hfill \\
  {d_k} = {d_{k - 1}} - {v_{k - 1}}\Delta T\cos {\theta _{k - 1}} + {\omega _{d,k}}, \hfill \\
  {v_k} = {v_{k - 1}} + {\omega _{v,k}}, \hfill \\
  {b_k} = {b_{k - 1}}\left( {1 + d_{k - 1}^{ - 1}{v_{k - 1}}\Delta T\cos {\theta _{k - 1}}} \right) + {\omega _{b,k}}, \hfill \\ 
\end{gathered}  \right. \hfill \\ 
\end{gathered}
\end{equation}
where ${{\bm{\omega }}_k} = \left[\omega_{\theta,k},\omega_{d,k},\omega_{v,k},\omega_{b,k}\right]^T$ represents the state noise, and $\Delta_T$ is the duration of one epoch.

With both models (\ref{measurement2}) and (\ref{state_model}) above, the RSU can predict and estimate the vehicle's state at each epoch via various approaches, e.g., Kalman filtering and factor graph based message passing algorithms \cite{Liu20,Yuan20Bayesian,Liu20tutorial}. The predicted angle is then employed to design the JCR beamformer for the next epoch. At the $k$th epoch, the received signal at the vehicle can be written as
\begin{equation*}
    {{\mathbf{y}}_C}\left( t \right) = {\beta _k}{{\mathbf{a}}^T}\left( {{M_T},{\theta _k}} \right){{\mathbf{a}}^*}\left( {{M_T},{{\hat \theta }_{k\left| {k - 1} \right.}}} \right)s_k\left( t \right) + {z_C}\left( t \right),
\end{equation*}
where $\beta_k$ is the path-loss of the LoS path, and ${z_C}\left( t \right)$ is the noise with variance $\sigma_C^2$. Accordingly, the achievable rate is obtained as
\begin{equation*} 
    {R_k} = \log \left( {1 + {{{{\left| {{\beta _k}{{\mathbf{a}}^T}\left( {{M_T},{\theta _k}} \right){{\mathbf{a}}^*}\left( {{M_T},{{\hat \theta }_{k\left| {k - 1} \right.}}} \right)} \right|}^2}{p_k}}}/{{\sigma _C^2}}} \right),
\end{equation*}
where $p_k$ is the power of the data stream $s_k\left(t\right)$. It can be seen that the achievable rate relies critically on the sensing and prediction accuracy of the AoA. Once the next AoA is accurately predicted, the RSU can keep tracking the vehicle while offering high-quality communication service when it is within the RSU's coverage. 

\emph{Remark:} We conclude here the superiorities of the JCR based predictive BF over conventional communication-only approaches, i.e., beam training and tracking:
\begin{itemize}
    \item First of all, JCR signalling removes the necessity of dedicated downlink pilots, as the whole JCR downlink block is exploited both for beam sensing and communication. This reduces the downlink overhead.
    \item Secondly, the uplink feedback is not required, since the RSU estimates the angle from the returned target echo signal instead of from the feedback, which reduces the uplink overhead.
    \item Thirdly, the quantization error generated in the uplink feedback is avoided. As such, the estimation of the vehicle's state can be performed in a continuous manner.
    \item Finally, JCR signalling achieves higher matched-filtering gain than that of beam training and tracking approaches, as the whole downlink frame, rather than a part of it, is tailored for both downlink sensing and data transmission.
\end{itemize}

\section{JCR: Radar-Centric Design}\label{sec-rcd}

Radar systems, particularly military radar, have the extraordinary capability of long-range operation, up to hundreds of kilometers. Therefore, a major advantage of implementing communication in radar systems is the possibility of achieving very long range communications, with much lower latency compared to satellite communications. However, the achievable data rates for such systems are typically limited, due to the inherent limitation in the radar waveform \cite{RN13, RN15,Hassanien16, Hassanien_2019}. 

Research on radar-centric JCR has been mainly focused on the information embedding technologies, and there are only limited works on other aspects such as communication protocol and receiver design based on the radar-centric JCR signals. In this section, we concentrate on more recent DFRC systems based on MIMO-OFDM, CAESAR and FH-MIMO radar, because of the remarkable benefits they can offer as described in Section \ref{sec-mimoradar}. 

\subsection{Embedding Information in Radar Waveform}

Realization of communication in radar systems needs to be based on either pulsed or continuous-wave radar signals. Hence information embedding with little interference on radar operation is one of the major challenges. This topic has been widely investigated, as reviewed in, e.g., \cite{Hassanien16, 8828016,Hassanien_2019}. Here, we summarize these techniques in Table \ref{tab-infoembed}. 

\begin{table*}
\caption{Summary of information embedding methods in radar-centric DFRC systems.} 
	\centering
			\begin{tabular}{|m{0.2cm}|m{1.2cm}|m{3.5cm}|m{5.5cm}|m{5.5cm}|}
		\hline
		\multicolumn{2}{|c|}{\textbf{Modulations}} & \multicolumn{1}{c|}{\textbf{Methods}} & \multicolumn{1}{c|}{\textbf{Advantages}} & \multicolumn{1}{c|}{\textbf{Disadvantages}} \\
		\hline
		\multirow{3}{*}{\rotatebox[origin=c]{90}{\parbox[c]{30mm}{\centering Modified Waveforms}}} 
		&Time-frequency Embedding & Apply various combinations of amplitude,
		phase and/or frequency shift keying to radar
		chirp signals \cite{4268440,Hassanien16, 8828016}, or map data to multiple chirp
		sub-carriers via the use of fractional Fourier Transform \cite{7485314}.
        & \begin{itemize}
		\item The chirp signal form remains when the inter pulse modulation is used, which is prefered in many radar applications.
		\item The waveform can be implemented in many existing radar systems with only modifications to the software.
		\end{itemize} 
        & \begin{itemize} 
        \item The slow time coding is restricted by the PRF of the radar, thereby limiting the maximum rate of communication.
        \end{itemize} 
        \\ \cline{2-5}
        
		& Code-domain Embedding & Modulate binary/poly-phased codes in radar signals using
		direct spread spectrum sequences \cite{4753277}. 
		& \begin{itemize}
		\item Naturally coexist with the CDMA / DSSS communication signal form.
		\item Enables covert communication by spreading the signal over the bandwidth of radar.
		\end{itemize}
		& \begin{itemize}
		\item Phase modulation will inevitably lead to spectrum alteration of the radar waveform, which may result in energy leakage outside the assigned bandwidth
		\end{itemize}
		\\ \cline{2-5}
		
		&Spatial embedding & Modulate information bits to the sidelobes of the radar beampattern \cite{Hassanien16, 8828016}. 
		& \begin{itemize}
			\item Has little impact on the radar sensing performance in the mainlobe.
		\end{itemize} 
	    & \begin{itemize}
	        \item The performance is sentive to the accuracy of array calibration and BF
	        \item The multi-path of radar signal may incur interference to the communication.
          \end{itemize}
	    \\ 
	    \hline
		 
		\multicolumn{2}{|m{1.4cm}|}{Index Modulation (No waveform modification) }	
		& Represent information by the indexes of antennas, frequencies, and/or codes of the signals \cite{Hassanien_2019,Huang20,Ma20,Wu20waveform}. 
		&  \begin{itemize}
			\item Naturally coexist with the radar functionality, with negligible impact on radar performance;
			\item Generally achieve higher data rates compared to modulation with modified waveform.
		\end{itemize}  
	   & \begin{itemize}
		\item Demodulation may be complicated;
		\item Demodulation performance is sensitive to channel if IM is applied to spatial domain;
		\item Codebook design could be a challenge.
	\end{itemize} \\	
		\hline
	\end{tabular}	
	\label{tab-infoembed}
\end{table*}

One of the particular techniques of interest is index modulation (IM). IM embeds information to different combinations of radar signals' parameters, over one or more domains of space, time, frequency and code \cite{xu2020dfrc, Hassanien_2019,Ma20}. Thus IM does not change the basic radar waveform and signal structure, and has negligible influence on radar operation. For MIMO-OFDM, CAESAR and FH-MIMO radar, IM can be realized via frequency selection/combination and/or antenna selection/permutation \cite{BouDaher16,Huang20IM,Baxter18, Wu20waveform}. Frequency combination selects different sets of frequencies, and antenna permutation allocates the selected frequencies to different antennas. Information is represented by the combinations and permutations. Let the number of combinations and permutations be $N_c$ and $N_p$, respectively. Then the number of bits can be represented is $\log_2N_c$ and $\log_2N_p$, respectively. Mathematically, frequency combination and antenna permutation can generally be combined. However, decoupling them is consistent with the way that the information is demodulated, as will be discussed later.  

For MIMO-OFDM radar with orthogonal frequency allocation, frequency combination allocates the total $N$ subcarriers to $M_T$ groups without repetition, with each group having at least one subcarrier \cite{BouDaher16,Wang18}. If without additional constraint on subcarrier allocation, there are a total of $N_c=C^N_{M_T}M_T^{N-M_T}$ combinations (i.e., Selecting $M_T$ out of $N$ subcarriers first to ensure each group to have at least one, and then the remained $N-M_T$ subcarriers can go to any of the $M_T$ groups); if each antenna needs to have the same number of $L_s=N/M_T$ subcarriers, the total number of combinations is $N_c=C^N_{L_s}C^{N-L_s}_{L_s}\ldots C^{L_s}_{L_s}=N!/(L_s!)^{M_T}$. The number of permutations of allocating $M_T$ groups of subcarriers to $M_T$ antennas is $N_p=M_T!$.

The DFRC system extended from CAESAR is proposed in \cite{Huang20IM}, where each virtual subarray is assumed to have the same number of antennas and use one frequency. Hence frequency combination selects $S$ out of $N$ frequencies, and $N_c=C^N_S$. The number of total antenna permutations is shown to be $N_p=M_T!/((M_T/S)!)^S$. 

\begin{figure}[t]
	\centering
	\includegraphics[width=0.8\linewidth]{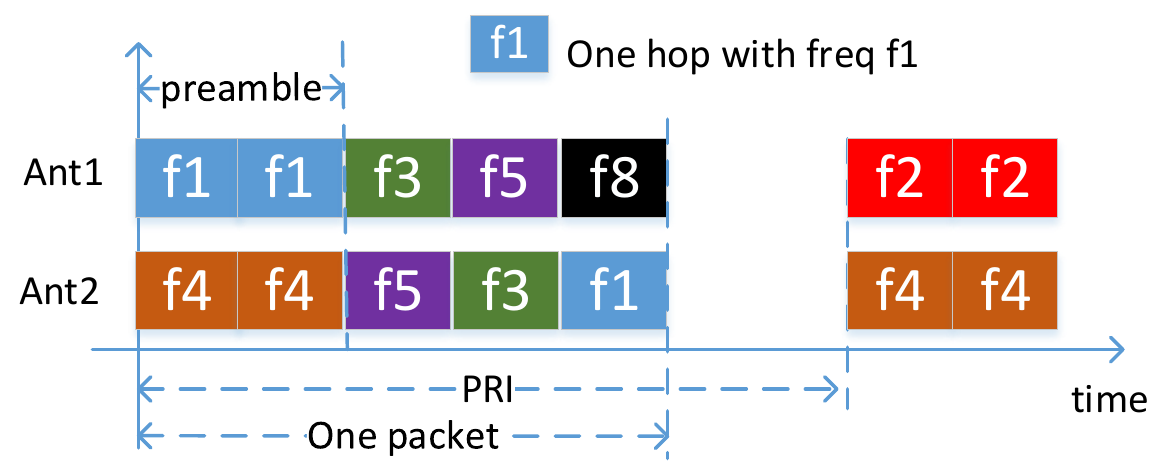}
	\caption{A simple example showing the packet structure of FH-MIMO DFRC with $N=8$ and $M_T=2$, consisting of a preamble with two identical hops and 3 hops with embedded information. A total of 4 bits can be conveyed in each hop with IM.}
	\label{fig-fhdfrc}
\end{figure}

For FH-MIMO DFRC systems \cite{Baxter18, wu2020integrating}, the total number of frequency combinations and antenna permutations are $N_c=C^N_{M_T}$ and $N_p=M_T!$, respectively. This corresponds to $S=M_T$ in \cite{Huang20IM}. Let $\Hfps_k$ denote the $M_T\times M_T$ antenna permutation matrix at hop $k$, which has only a single non-zero element, $1$, in each row and column. The transmitted signal of the FH-MIMO DFRC can be represented as
\begin{align}
\xt_{R}(t)&=\Hfps_k \bm{\psi}(t).
\end{align}
Note that both $\Hfps_k$ and the frequency set $\mathcal{F}_k$ vary with $k$ and are determined by the information bits. One simple example of information embedded FH-MIMO DFRC for packet communication is shown in Fig. \ref{fig-fhdfrc}.

It is noted that the values of $N_c$ and $N_p$ above define the maximum achievable bit rates only, without considering the communication reception performance. In practice, the number of actually used combinations and permutations may be reduced, due to the overall system design and the consideration of the demodulation complexity and performance. In particular, demodulating the bits embedded in antenna permutation is much more difficult and subject to higher demodulation error, compared to demodulating those embedded in frequency combination. These issues will be discussed in detail in Section \ref{sec-recep} and \ref{sec-codebook}. In addition, the impacts of information embedding on radar performance should also be evaluated, such as the ambiguity functions in the time domain \cite{Wu20waveform} and in the angular domain \cite{Wang18},. 

\subsection{Signal Reception and Processing for Communications}\label{sec-recep}
Since IM in all the three systems involve FH, and their receiver processing methods are similar in many aspects, we use FH-MIMO DFRC as an example to illustrate the signal reception and processing for its relative simplicity. Overall, the research on the receiver design for these systems are still limited. Our overview here is mainly based on \cite{Baxter20, Wu20waveform,wu2020integrating}, and also incorporates works on MIMO-OFDM and CAESAR DFRCs \cite{Wang18, Huang20IM} into the framework of FH-MIMO DFRC.

Consider a receiver with $M_R$ antennas. The signal received from each antenna is passed to a mixer with local oscillator frequency $f_c$, which is generally the central frequency of the $N$ subbands. Assume narrowband communications and the difference between multipath delays $|\tau_\ell-\tau_{\ell'}|\ll T$. Referring to \eqref{eq-Ht1} and \eqref{eq-ytnb} and ignoring the variation of Doppler frequencies, the noise-free baseband received signal can be approximated as
\begin{align}
\yt(t)&=\sum_{m=1}^{M_T}\sum_{\ell=1}^L c_\ell e^{j2\pi (f_{k,m}-f_c)(t-\tau-\tau_o(t))}\\ &\quad\av(M_R,\phi_\ell)\av^T(M_{T},\theta_{\ell})\hfps_{k,m}g(t-\tau-\tau_o(t)-kT)\notag,
\end{align}
where $\hfps_{k,m}$ is the $m$-th column of $\Hfps_k$, $c_\ell\in\mathbb{C}$ is the equivalent path coefficient, subsuming multiple terms, and $\tau_\ell\approx \tau$ is used. The baseband signal is then sampled at $T_s=1/B$, generating $L_p=\lfloor T/T_s\rfloor$ samples per hop. Let
\begin{align}
\Ht\triangleq\sum_{\ell=1}^L c_\ell \av(M_R,\phi_\ell)\av^T(M_{T},\theta_{\ell}).
\end{align}
For the simplicity of presentation, assume that $g(t)$ is a rectangular windowing function. Assume that synchronization is done perfectly. We can stack $L_p$ measurements from all $M_R$ antennas to a matrix $\Yt_k$, which is given by
\begin{align}
\Yt_k=\Ht\bm{B}_k\bm{\Phi}^T=\sum_{m=1}^{M_T} \Ht \hfps_{k,m} \bm{\psi}_{k,m}^T,
\label{eq-ytim}
\end{align}
where $\bm{\Phi}=[\bm{\psi}_{k,1},\ldots,\bm{\psi}_{k,M_T}]$, $\hfps_{k,m}$ is the $m$-th column of $\Hfps_k$, and $\bm{\psi}_{k,m}=\{e^{j2\pi (f_{k,m}-f_c)\ell_p}\}, \ell_p=0, \ldots, L_p-1$. 

\subsubsection{Demodulation}
The task of demodulation is to retrieve information bits from $\Yt_k$. We assume perfect synchronization and channel estimation here. It will become clear that channel estimation may not be necessary if only frequency combination needs to be identified.

An optimal formulation of the demodulator can be based on the maximum likelihood principle \cite{Huang20IM}. But it has very high computational complexity and is infeasible for practical implementation. Alternatively, we can apply sub-optimal methods such as CS techniques.  

The patterns of both frequency combination and antenna permutation can be identified by formulating a sparse recovery problem. The basic idea is to construct an $L_p\times N$ dictionary matrix $\bm{\Phi}_d$ by expanding $\bm{\Phi}$ to cover all $N$ subband frequencies. Each column of $\bm{\Phi}_d$ has the similar expression with $\bm{\psi}_{k,m}$ for each subband frequency. Then we can get a 1-D MMV-CS formulation as
\begin{align}
\Yt_k^T=\bm{\Phi}_d \A_\text{MMV},
\end{align}
where only $M_T$ out of $N$ rows in $\A_\text{MMV}$ are non-zero, corresponding to $(\Ht\bm{B}_k)^T$. The $M_T$ frequency estimates, which correspond to the frequency combination pattern, can then be found by using one of the well-known MMV-CS recovery algorithms. The estimate of the $M_T$ non-zero rows of $\A_\text{MMV}$, obtained in the recovery process, can be used to find the antenna permutation pattern, by matching them with $(\Ht\bm{B}_k)^T$. This matching process can be realized in either a simpler row-wise way or a more complicated process jointly across all rows.

Another simpler sub-optimal method is to exploit the orthogonality of frequencies across antennas and apply a discrete Fourier transform (DFT) matrix $\F$:
\begin{align}
\Yt_k\F=\Ht\bm{B}_k\bm{\Phi}_k^T\F,
\end{align}  
where each row of $\bm{\Phi}_k^T\F$ is the windowed DFT output of a single tone signal and its waveform has the shape of an impulse with a single peak. Thus each row of $\Yt_k\F$ represents the weighted sum of these impulses. Therefore, the frequency combination pattern may be identified via locating the peaks. This is particularly effective when either the inverse of $\Ht$ exists or when the LOS path is dominating in $\Ht$. In the former, we can compute $\Ht^{-1}\Yt_k$ and obtain $\bm{B}_k\bm{\Phi}_k^T\F$. This leads to simple identification of both frequency combination and antenna permutation patterns, as $\bm{B}_k$ is a permutation matrix. In the latter, the frequency combination pattern can be found via the peaks and the antenna permutation pattern is determined via exhaustive searching, even in a single antenna receiver \cite{Wu20waveform,wu2020integrating}.

\subsubsection{Channel Estimation}
An accurate estimate of $\Ht$ is critical for demodulation. However, it is challenging to design and incorporate long training sequences, which is essential for estimating $\Ht$, in FH-MIMO DFRC systems. This is because training sequence requires certainty, which will affect the randomness of FH radar operation. 
 
There are very limited results on channel estimation for FH-DFRC systems with IM. In \cite{Wu20waveform}, both synchronization and channel estimation are investigated for a single antenna receiver, with the consideration of packet communications. For channels with a dominating LOS-path, which could be a typical operating condition for radar-centric DFRC, a frame structure is proposed with two identical hops serving as preamble followed by hops with embedded information. The two identical hops are designed to enable effective estimation of timing offset, carrier frequency offset and channel. To simplify synchronization and channel estimation, re-ordered hopping frequencies are used, which slightly reduces the information embedding capability in terms of antenna permutation. Timing offset and channel estimators are proposed by exploiting the signal differences between two neighbouring antennas. The work is also extended to NLOS channels by using incomplete sampled hops and judiciously designed hopping frequencies to combat inter-hop and inter-antenna interference.

\subsection{Codebook Design}\label{sec-codebook}
The codebook determines how the patterns of frequency combination and antenna permutation are selected and mapped to information bits. As was disclosed in \cite{Huang20IM}, the achievable communication rates are largely constrained by antenna permutation, as its demodulation performance is sensitive to the differences between different columns of $\Ht$.

The design criterion can hence be formulated based on the distance between two codewords of antenna permutation:
\begin{align}
\lambda(m,m')=\parallel\Ht \hfps_{k,m}-\Ht \hfps_{k,m'}\parallel_2^2.
\label{eq-lambdamm}
\end{align}
Maximizing the minimal distance among all $\lambda(m,m')$ is a typical design criterion. Since directly searching via \eqref{eq-lambdamm} is computational complicated, a method of projection into a lower dimensional plane is proposed in \cite{Huang20IM}. However, given that the design needs to be updated once $\Ht$ is changed, the complexity is still very high.

Such a complicated design may be avoided by using pre-compensation. For example, for LOS-path dominating channels, the channel differences between antenna permutations will be small when the AoD is small. In \cite{wu2020integrating}, an element-wise phase compensation method is proposed to remove the AoD dependence of demodulating antenna permutation. Thus the distances between different codewords become identical.

In addition to its impact on communication performance, codebook may also affect the radar performance, for example, the ambiguity function as evaluated in \cite{Baxter20, Wu20waveform}. More specifically, it is demonstrated in \cite{Baxter20} that the probability of radar waveform degeneration can be reduced by spreading the available frequency hops between waveforms as evenly as possible; and in \cite{Wu20waveform}, it is shown that by constraining the codewords, the receiver processing can be largely simplified, with negligible impact on the radar ambiguity function.

\section{JCR: Joint Design and Optimization} \label{sec-waveform}

Although no clear boundary exists between the third category of JCR and the other two, there is more freedom here in terms of signal and system design. That is, JCR technologies can be developed without being constrained to existing C\&R systems, and they can be designed and optimized to balance the requirements for C\&R, potentially providing a better trade-off between the two functions.

Joint waveform optimization is a key research problem here. It can be conducted in multiple domains, using various performance metrics jointly for C\&R. In this section, we review several typical DFRC waveform optimization schemes. For simplicity, we will mainly consider underlying signals based on narrowband single carrier communications, and the extension to OFDM signals is generally straightforward.

To avoid confusion, we use $\Ht_C$ and $\Ht_R$ for communication and radar channels, respectively; and for simplicity, we assume $M_T=M_R=M$. Note that $\Ht_C\neq\Ht_R$ in downlink sensing and $\Ht_C=\Ht_R$ in uplink sensing, as detailed for PMNs in \cite{8827589,Zhangpmn20}. Referring to \eqref{eq-ytnb}, collect $K$ received signal vectors over a CPI and stake them to a matrix, generating $\Yt_C=\Ht_C\Xt +\Zt_C$ for communications, and $\Yt_R=\Ht_R\Xt +\Zt_R$ for sensing. We assume the AWGN matrices $\Zt_R$ and $\Zt_C$ both have zero mean and element-wise variance $\sigma_z^2$.

\subsection{Waveform Optimization via Spatial Precoding}

When different beams are needed for C\&R, the precoding matrix $\Pc$ can be optimized. The basic optimization formulation is as follows:
\begin{align}
&\arg\max_\Pc \lambda(\Pc), \quad\text{s.t. Constraints 1, 2} \cdots,
\label{eq-wfopt}
\end{align}
where $\lambda(\Pc)$ is the objective function. There could be various methods and combinations in defining the objective functions and the constraints. Each can be either for communication or sensing individually, or a weighted joint function. Next, we review three types of optimizations that consider mutual information (MI), waveform mismatch, and estimation accuracy for the sensing performance, respectively.

\subsubsection{Mutual Information (MI) Based} 

MI is well known for communication systems, and the usage of MI for radar waveform design can also be traced back to 1990s \cite{Bell1993Information}. MI for radar sensing measures how much information about the channel, the propagation environment, is conveyed to the receiver. The conditional MI is defined as the entropy between the sensing channels and the received signals, conditional on the transmitted signal. Mathematically, this can be represented as \cite{yuan2020waveform}
\begin{align*}
I_R(\Ht_R; \Yt_R|\Xt)=M\log_2\left(\text{det}\left(\frac{1}{\sigma_z^2}\Xt^H\bm{\Sigma}_{H_R}\Xt+\I_K\right)\right),
\end{align*}
where $\I_K$ is an identity matrix of size $K\times K$, and $\bm{\Sigma}_{H_R}=\text{E}[\Ht_R\Ht_R^H]/M$. The conditional MI for communication is given by
\begin{align*}
I_C(\Xt; \Yt_C|\Ht_C)=K\log_2\left(\text{det}\left(\frac{1}{\sigma_z^2}\Ht_C^H\bm{\Sigma}_{X}\Ht_C+\I_M\right)\right),
\end{align*}
where $\bm{\Sigma}_{X}=\text{E}[\Xt\Xt^H]/K$.

Based on the two MI expressions, we can formulate various optimization problems. In \cite{RN15}, the estimation rate, defined as the MI within a unit time, is used for analyzing the radar performance, together with the capacity metric for communications. In \cite{yliu17}, a weighted sum of MI for both C\&R is formulated as the objective function of optimization for a single-antenna OFDM DFRC system. In \cite{yuan2020waveform}, a more complicated MI-based joint optimization is conducted for MIMO DFRC systems, by taking into consideration practical packet structure with orthogonal training sequences and random data symbols, and the channel estimation error. 

A general and flexible formulation can be based on a weighted sum of the two MIs \cite{yliu17, yuan2020waveform}
\begin{align}
F=\frac{w_R}{F_R}I_R(\Ht_R; \Yt_R|\Xt)+\frac{1-w_R}{F_C}I_C(\Xt; \Yt_C|\Ht_C),
\label{eq-miopt}
\end{align}
where $F_C$ and $F_R$ are the maximal MI for C\&R individually, and are treated as two known constants in the optimization, and $w_R\in[0,1]$ is a weighting factor. The function in \eqref{eq-miopt} is concave, and can be maximized by using, e.g., the Karush-Kuhn-Tucker (KKT) conditions. The optimal $\Pc$ turns out to be a water-filling type of solution that jointly considers the distributions of the  eigen-values of $\bm{\Sigma}_{X}$ and $\bm{\Sigma}_{H_R}$.

\subsubsection{Waveform/Beampattern Similarity Based} 
The DFRC waveform is typically expected to possess some useful properties that are beneficial for radar sensing, e.g.,  good auto- and cross-correlations, high peak-to-sidelobe level ratio (PSLR), low peak-to-average power ratio (PAPR), and resilience to clutter and interference. Nevertheless, it could be quite challenging to implement all these features simultaneously in a single waveform, especially in the case of JCR, where the randomness in the communication data and channels may break down the structure of the waveform tailored for sensing. 

To cope with this issue, one may consider to optimize the DFRC waveform/beampattern while approximating a well-designed benchmark radar signal, which is known to have the desired characteristics above. This could be achieved by imposing a figure-of-merit for the waveform/beampattern similarity, either in the objective function or in the constraints. As an example, the JCR BF design in \cite{Liu18mumimo} aims to approximate a baseline radar beampattern while guaranteeing the individual signal-to-interference-and-noise ratio (SINR) $\gamma_k$ for $K$ single-antenna downlink users. Accordingly, the optimization problem can be formulated as
\begin{equation}\label{beampattern_approx}
\begin{gathered}
  \mathop {\min }\limits_{{\mathbf{\Pc}},\beta } \;\;\left\| {{\mathbf{P}}{{\mathbf{P}}^H} - \beta {\mathbf{R}}} \right\|_F^2 \hfill \\
  s.t.\;\;\;\;\beta  \ge 0, \quad {\gamma _k} \ge {\Gamma _k},\forall k, \hfill \\
  \;\;\;\;\;\;\;\;\;\operatorname{diag} \left( {{\mathbf{P}}{{\mathbf{P}}^H}} \right) = \frac{{{P_T}}}{{{M_T}}}{{\mathbf{1}}_{{M_T}}}, \hfill \\
\end{gathered}
\end{equation}
where $\mathbf{P}\in\mathbb{C}^{M_T \times K}$ is the DFRC BF/precoding matrix to be designed, $\mathbf{R}\in\mathbb{C}^{M_T \times M_T}$ represents the spatial covariance matrix for a benchmark radar waveform, which generates a favorable MIMO radar beampattern $P_d\left(\theta\right)$, given by
\begin{equation}
    {P_d}\left( \theta  \right) = {{\mathbf{a}}^H}\left( {{M_T},\theta } \right){\mathbf{Ra}}\left( {{M_T},\theta } \right).
\end{equation}
It can be readily observed that the cost function in (\ref{beampattern_approx}) is the Euclidean distance between covaraince matrices for the DFRC waveform and its pure radar benchmark. Moreover, $\beta \ge 0$ is a scaling factor. The remaining constraints aim to ensure the SINR for each user, as well as to restrict the per-antenna transmit power, with $P_T$ being the overall power budget available. While problem (\ref{beampattern_approx}) is non-convex, it can be sub-optimally solved via the semidefinite relaxation (SDR) approach or manifold based algorithms \cite{Liu18mumimo}.

In addition to the above design that approximates the MIMO radar beampattern, a more straightforward method is to directly approximate the MIMO radar waveform itself. Under the same $K$-user MU-MIMO downlink scenario, let us first rewrite the MIMO communication signal model (\ref{eq-ytnb}) in a discrete matrix form as
\begin{equation}\label{MUI_model}
{\mathbf{Y}}_C = {{\mathbf{H}}_C\mathbf{X}} + {\mathbf{Z}}_C = {\mathbf{S}} + \underbrace {\left( {{{\mathbf{H}}_C\mathbf{X}} - {\mathbf{S}}} \right)}_{{\text{MUI}}} + {\mathbf{Z}}_C,
\end{equation}
where $\mathbf{S} \in \mathbb{C}^{K \times L}$ contains the information symbols intended for $K$ communication users. The second term at the right-hand side of the second equality is known to be the multi-user interference (MUI). If the MUI is minimized to zero, then (\ref{MUI_model}) boils down to an AWGN transmission model, where the fading effect incurred by the channel $\mathbf{H}_C$ vanishes. By noting this fact, \cite{Liu18} formulates the following optimization problem to design a DFRC waveform matrix $\mathbf{X}$
\begin{equation} \label{CM_JCR}
\begin{gathered}
  \mathop {\min }\limits_{\mathbf{X}} \;\;\left\| {{{\mathbf{H}}_C\mathbf{X}} - {\mathbf{S}}} \right\|_F^2 \hfill \\
  s.t.\;\;\;\;{\left\| {\operatorname{vec} \left( {\mathbf{X}} \right) - \operatorname{vec} \left( {{{\mathbf{X}}_0}} \right)} \right\|_\infty } \le \varepsilon , \hfill \\
  \;\;\;\;\;\;\;\;\;{\left| {{x_{i,j}}} \right|^2} = \frac{{{P_T}}}{{{M_T}}},\forall i,j, \hfill \\ 
\end{gathered}
\end{equation}
where the first constraint is to explicitly control the distance between $\mathbf{X}$ and the benchmark $\mathbf{X}_0$ in an $L_\infty$-norm sense, with a given similarity coefficient $\varepsilon$. The second constraint, on the other hand, requires $\mathbf{X}$ to be constant-modulus (CM), i.e., with a 0dB PAPR. $\mathbf{X}_0$ could be any CM radar signal matrix, e.g., orthogonal chirp waveform. While problem (\ref{CM_JCR}) is again non-convex and NP-hard in general, a branch-and-bound (BnB) algorithm is developed in \cite{Liu18}, which finds its global optimum within only tens of iterations.

\subsubsection{Estimation Accuracy Based}  

The accuracy of sensing parameter estimation is important for radar sensing. Since the received signals are not a linear function of the sensing parameters, it is generally difficult to get closed-form expressions for, e.g., the mean square error (MSE) of the estimates, and to apply them in the optimization directly. Alternatively, we can derive and use the CRLBs of the estimates, which are low bounds of the MSE. The CRLBs of signal estimates can be derived via the inverse of the Fisher information matrix (FIM). For estimates based on radar signals, they are well known, e.g., as reported in \cite{6178073}. For communication signals, the CRLBs for some sensing parameters based on the beamspace channel models have also been derived, e.g., in \cite{Larsen09}. 

DFRC waveform optimization based on the CRLBs for sensing performance has also been studied in the literature. However, due to the complicated expressions of the CRLBs, it is generally challenging to obtain closed-form solutions in such optimizations. In \cite{LIU2017331}, considering a single-antenna OFDM DFRC system, Pareto-optimal waveform design approaches are proposed to simultaneously improve the estimation accuracy of range and velocity and the channel capacity for communications. The Pareto-optimal solutions are obtained for a multiobjective optimization problem, and cannot be improved with respect to any objective function without deteriorating other objective functions, and hence the solutions are suboptimal. In \cite{ni2020waveform}, waveform optimization is studied with the application and comparison of multiple sensing performance metrics including MI, MMSE and CRLB, together with an approximated SINR metric for communications. It is shown that there are close connections between MI-based and CRLB-based optimizations, and the MI-based method is more efficient and less complicated compared to the CRLB-based method. Specific to the CRLB-base method, a closed-form solution is obtained for some special cases, while an iterative algorithm is proposed as a general solution.

\subsection{Multibeam Optimization for Analog Array}\label{sec-multibeam}
Steerable analog array, which is also the basic component of a hybrid array, could be widely applied in mmWave JCR systems. A JCR system may need to support communication and sensing at different directions, which is challenging to address given the limits on the BF capability of analog array. A good solution is the multibeam technology \cite{8550811,Luo19jcas,Luo20multibeam}. The multibeam consists of a fixed subbeam dedicated to communication and a scanning subbeam with directions varying over different communication packets. By optimizing the beam with multiple subbeams, communication and radar sensing at different directions can be supported simultaneously with a single signal. This can largely extend the field of view of sensing, for example, in the 802.11ad DFRC with the single carrier PHY. The multibeam can be applied at both transmitter and receiver, while transmitter is considered here. 

Two classes of methods have been proposed for the multibeam optimization: the subbeam-combination method \cite{8550811,Luo19jcas,Luo20multibeam} via constructively combining two pre-generated subbeams according to given criteria, and the \textit{global optimization} \cite{Luo20multibeam} by jointly considering the C\&R BF requirements and optimizing a single BF vector directly. These methods can also be extended to full digital arrays.

\subsubsection{Subbeam Combination}

In the subbeam-combination method, two basic beams for C\&R are separately generated according to the desired BF waveform, using, e.g., the iterative least squares method \cite{8550811}. The two beams are further shifted to the desired directions by multiplying a sequence, and then combined by optimizing a power distribution factor $\rho$ and a phase shifting coefficient $\varphi$. 

The BF vector $\w_t$ in \eqref{eq-wt} can be represented as
\begin{align}\label{eq-wt}
\mathbf{w}_T=\sqrt{\rho}\w_{T,F}+\sqrt{1-\rho} e^{j\varphi}\w_{T,S},
\end{align}
where $\w_{T,F}$ and $\w_{T,S}$ are the BF vectors that are predetermined, corresponding to the fixed subbeam and scanning subbeam, respectively. The value of $\rho$ can typically be determined via balancing C\&R distances \cite{Luo19jcas}, and the optimization is conducted mainly with respect to $\varphi$, which can ensure the pre-generated subbeams are coherently combined when generating the multibeam.  

The multibeam optimization problem can then be formulated with desired objective functions and constraints. Consider one example of maximizing the received signal power for communications with constrained BF gain in scanning directions. Let the threshold of the BF gain in the $i$-th sensing direction $\phi_{i}$ be $C^2_{i}(1-\rho)M_T$, where $C_{s_i}\in[0,1]$ is a scaling coefficient, representing the ratio between the gain of the scanning subbeam in the interested direction and the maximum gain that the array can achieve, i.e., $(1-\rho)M_T$. We can formulate the constrained optimization problem as 

\begin{align}
& {\varphi}^{(1)}_\text{opt}=\arg\max_{\varphi}\dfrac{\w_T^H{\mathbf{H}_C}^H{\mathbf{H}_C}\w_T}{\|\w_T\|^2},\label{eq-stscan}\\
&\text{s.t. }\ \dfrac{|\av^{T}(M_T,\phi_{i})\w_T|^2}{||\w_T||^2}\geq C^2_{i} (1-\rho)M_T, \  i=1,2, \cdots, N_s, \notag
\end{align}
where $N_s$ is the number of the total constraints, $\phi_i$s are the angles of interest where the BF gain is constrained, and a maximal ratio combiner is assumed to be applied at the communication receiver.
	
With $\w_{T,F}$, $\w_{T,S}$ and $\rho$ being predetermined, the optimization problem can be solved by first finding the unconstrained optimal solution for the objective function and then looking for its intersection with the intervals determined by the constraints. Closed-form solutions can then be obtained as detailed in \cite{Luo20multibeam}.

The subbeam-combination method enables simple and flexible multibeam generation and optimization, and hence is promising for practical applications that require fast adaptation to changing BF requirements. It is particularly useful for mmwave systems where directional BF is used.  

\subsubsection{Global Optimization}

The subbeam-combination method is an efficient low-complexity solution, but it is suboptimal. A globally optimal solution can be obtained by solving a problem formulated directly with respect to $\w_T$. Considering a similar example of maximizing the received signal power for communication with constraints on BF waveform, the formulation can be represented as
\begin{subequations}
\begin{align} 
&\w_{t,\text{opt}}= \argmax_{\w_T, \w_T^H\w_T=1} \ \w_T^H{\mathbf{H}_C}^H{\mathbf{H}_C}\w_T, \label{eq-OutPow2}\\
&\text{s.t.}\quad\|\A(M_T,\phi_i)\w_T-\df_v)\|^2\leq \varepsilon_w, \ \text{and/or} \label{eq-con1}\\
&\qquad\ |\av^T(M_T,\phi_{i})^T\w_T|^2\geq \varepsilon_{i},\ i=1,2,\cdots, N_s,  \label{eq-con2}, 
\end{align}
\label{eq-global}
\end{subequations}
where \eqref{eq-con1} bounds the mismatches between the generated BF waveform and the desired one $\df_v$, \eqref{eq-con2} constrains the gain of the scanning subbeam in $N_s$ directions, and $\varepsilon_w$ and $\varepsilon_{i}$ are the thresholds for these constraints. These constraints can be applied individually or jointly.  

The optimization in \eqref{eq-global} is generally a nonconvex NP-hard problem. In \cite{Luo20multibeam}, this problem is converted to a homogeneous quadratically constrained quadratic program (QCQP), which is then solved by the SDR technique.

The global optimization method provides a benchmark for suboptimal multibeam optimization schemes. Simulation results in \cite{Luo20multibeam} demonstrate that there exist a loss up to 10\% in the received signal power for communications and the BF waveform, when the subbeam combination method is applied.

\subsection{Signal Optimization in Other Domains}

In addition to spatial optimization, communication signals can also be optimized across the time and frequency domains, to improve the estimation accuracy of sensing parameters. For example, non-uniform preamble is proposed to improve the Doppler estimation performance in \cite{Kumari20}; a modified Golay complementary sequence is proposed for 802.11ad-based JCR systems in \cite{Duggal20} to reduce the sidelobe and achieve improved ranging and Doppler estimation; and the idea of using and optimizing nonequidistant subcarriers in MIMO-OFDM radar in \cite{Hakobyan20} can also be extended to a JCR MIMO-OFDM system. Here, we briefly illustrate the idea by referring to the work in \cite{Kumari20}.

In \cite{Kumari20}, for a single data-stream single-carrier JCR system based on 802.11ad, non-uniformly placed preambles are proposed to enhance velocity estimation accuracy. The signal has a packet structure, consisting preamble and data payload. Radar sensing uses the preamble only. A novel metric of distortion MMSE (DMMSE) is developed for communications in \cite{Kumari20}. A log-scale is then applied to the DMMSE and the CRLB for radar, to achieve proportional fairness between C\&R, such that the two log-scaled metrics can directly be added up in the optimization objective function. The objective function is given by
\begin{align*}
w_C\frac{1}{K}\text{Tr}[\log_2\text{DMMSE}] +w_R\text{Conv}(\frac{1}{L_v}\text{Tr}[\log_2\text{CRLB}]),
\end{align*}
where $K$ is the number of symbols, $L_v$ is the number of velocities to be estimated, $\text{Tr}[\cdot]$ denotes the trace of a matrix, $\text{Conv}(\cdot)$ denotes the convex hull operation, and $w_C$ and $w_R$ are weighting factors for C\&R, respectively.

Based on the objective function and some constraints, the number and position of preambles are optimized in \cite{Kumari20}. It is found that when preambles are equally spaced, the performance of radar or communications cannot be effectively improved without affecting the other. Comparatively, non-uniform preambles are found to achieve a better performance trade-off between C\&R, particularly at large radar distances.   

\section{Conclusion}\label{sec-final}
JCR is a promising technology that can be used to revolutionize both traditional communication and radar systems. Although the potentials and prospects for integrating C\&R are great, there are many challenges and open research problems to be addressed due to the inherent differences of the signal formats, and system and network structures between them. Signal processing techniques are key enablers to make the integration happen. For communication-centric JCR, accurate and practical sensing parameter estimation algorithms are the key, and a viable solution to the clock asynchrony problem can relax the full-duplex requirement. Sensing assisted communications is a great way of exploiting the benefit of integration. For radar-centric JCR, how to improve the communication rates without notable impact on radar operation is a big challenge. While it is shown that applying index modulation to frequency-hopping DFRC is an attractive solution, more signal processing techniques are needed for improving the communication receiver. For joint design and optimization, the journey is just started, with most work being focused on waveform optimization in existing systems. A fresh look at the joint requirements for C\&R may lead to more efficient solutions, particularly those underlying frequency hopping and millimeter wave signals that have excellent potentials for both high data rates and radar resolution.

\bibliographystyle{IEEEtran}
\bibliography{IEEEabrv,references_updated}
 \end{document}